\documentclass{ws-procs975x65}
\usepackage{amssymb,epsfig,amsmath,euscript,array}

\newcommand{\be}{\begin{equation}}
\newcommand{\ee}{\end{equation}}

\def\bd{\begin{document}}
\def\ed{\end{document}}
\def\nn{\nonumber}
\def\bea{\begin{eqnarray}}
\def\eea{\end{eqnarray}}
\let\bm=\bibitem
\let\la=\label

\newcommand{\EQ}[1]{\begin{equation} #1 \end{equation}}
\newcommand{\AL}[1]{\begin{subequations}\begin{align} #1 \end{align}\end{subequations}}
\newcommand{\SP}[1]{\begin{equation}\begin{split} #1 \end{split}\end{equation}}
\newcommand{\ALAT}[2]{\begin{subequations}\begin{alignat}{#1} #2 \end{alignat}\end{subequations}}
\def\beqa{\begin{eqnarray}}
\def\eeqa{\end{eqnarray}}
\def\beq{\begin{equation}}
\def\eeq{\end{equation}}

\def\hf{{\textstyle{1\over2}}}
\def\sst{\scriptscriptstyle}
\def\L{\Lambda}
\def\l{\lambda}
\newcommand{\tr}{\mbox{tr}}
\newcommand{\slsh}[1]{/ \!\!\!\! #1}

\def\vev#1{\langle{#1}\rangle}
\begin{document}

\title{Gauge Theory Amplitudes, Scalar Graphs and Twistor Space}

\author{Valentin V. Khoze}

\address{
Department of Physics and IPPP\\
University of Durham\\
Durham, DH1 3LE\\
United Kingdom\\
E-mail: valya.khoze@durham.ac.uk }

\maketitle

\bigskip
\bigskip

\abstracts{
We discuss a remarkable new approach initiated by Cachazo, Svrcek and Witten
for calculating gauge theory amplitudes.
The formalism amounts to an effective scalar perturbation theory which
in many cases
offers a much simpler alternative to the usual Feynman diagrams
for deriving $n$-point amplitudes in gauge theory.
At tree level the formalism works
in a generic gauge theory, with or without supersymmetry,
and for a finite number of colours. There is also a growing evidence that
the formalism works for loop amplitudes.
}

\bigskip
\bigskip
\bigskip
\bigskip
\bigskip
\bigskip
\bigskip
\bigskip

\abstracts{ To appear
in Ian Kogan Memorial Volume ``From Fields to Strings:
Circumnavigating Theoretical Physics''}

\newpage

\tableofcontents

\newpage

\section{Introduction}

In a recent paper \cite{CSW} Cachazo, Svrcek and Witten (CSW)
proposed a new approach for calculating
scattering amplitudes of $n$ gluons.
In this approach tree amplitudes in gauge theory
are found by summing tree-level scalar diagrams.
The CSW formalism \cite{CSW} is constructed in terms of
scalar propagators, $1/q^2,$ and
tree-level
maximal helicity violating (MHV) amplitudes,
which are interpreted as new scalar vertices.
The MHV vertices already contain an arbitrary number of gluon lines,
and are know explicitly \cite{PT,BG}.
Using multi-particle MHV amplitudes as effective vertices in a new perturbation theory
enables one to save dramatically on a number of permutations in usual
Feynman diagrams.

This novel diagrammatic approach \cite{CSW} follows from an
earlier construction \cite{Witten}
of Witten  which related
perturbative amplitudes of conformal ${\mathcal  N}=4$ supersymmetric gauge theory
in the large $N_{\rm c}$ limit
to D-instanton contributions in a topological string theory
in twistor space. The key observation of \cite{Witten,CSW} is that tree-level
and also loop diagrams in SYM posses a tractable geometric structure
when they are transformed from Minkowski to twistor space.

The results \cite{Witten,CSW} have been tested and further developed in gauge
theory in \cite{Zhu1,GK,Zhu2,BBK,Kosower,GGK,CSW2,BST},
and in string theory and supergravity in
\cite{RSV1,RSV2,BM,NV,W,GLMN,Siegel,Giombi,Popov,BW}.

The new perturbation theory involves scalar diagrams since
MHV vertices are scalar quantities. They are linked together
by scalar propagators at tree-level, and the internal lines are continued
off-shell in a particular fashion. The final result
for any particular amplitude can be shown to be
Lorentz-covariant and is independent of a particular choice for the off-shell
continuation. The authors of \cite{CSW} derived new expressions
for a class of tree amplitudes with three consecutive negative helicities
and any number of positive ones.
It has been verified already in \cite{CSW}
that the new scalar graph approach agrees with a number
of known standard results for scattering
amplitudes in pure gauge theory. Furthermore, it was shown in \cite{Zhu1}
that all $\overline{\rm MHV}$ (or googly) amplitudes -- i.e. amplitudes with
two positive helicity gluons and an arbitrary number of negative ones --
are reproduced correctly in the CSW formalism.
Recursive relations for constructing generic tree-level non-MHV amplitudes
in the CSW formalism were obtained in \cite{BBK}. Moreover,
general next-to-MHV gluonic amplitudes were derived in \cite{Kosower}
at tree level. These are the amplitudes where any three of $n$ gluons have
negative helicities.

We conclude that there is a sufficient evidence
that the CSW method works correctly and remarkably well at tree level and
for gluons only amplitudes.
Given this and also the fact that at present there is no detailed derivation
of the CSW rules neither in gauge theory, nor in string theory, we would like to see
how the method works in more general settings.
There is a number of questions one can ask
about the CSW formalism from the gauge theory perspective:
\begin{enumerate}
\item{} Does the CSW method work only in pure gauge sector at tree level or can
it be applied to supersymmetric theories?
\item{} If the method does apply to supersymmetric theories, does it work
in ${\mathcal  N}=1$ theory or in ${\mathcal  N}=4$ theory or in a generic supersymmetric Yang-Mills?
\item{} Does it work for diagrams with fundamental quarks in a non-supersymmetric
$SU(N)$ theory, i.e. in QCD?
\item{} Can we work with finite number of colours?
\item{} Can the CSW approach be used for practical calculations of amplitudes needed
in phenomenological applications?
\item{} Ask the five questions listed above for amplitudes at loop level.
\end{enumerate}
The goal of these notes is to discuss and answer some of these questions.

It is often said that any gauge theory at tree level behaves as if it was
supersymmetric.
More precisely, in a supersymmetric theory, the non-supersymmetric sector
and the superpartners are completely decoupled at tree level.
This is because at tree level
superpartners cannot propagate in loops.
This observation, on its own, does not answer the question of how to
relate amplitudes with quarks to amplitudes with gluinos.
The colour
structures of these amplitudes are clearly
different\footnote{Also, amplitudes with gluons and gluinos
are automatically planar at tree level. This is not the case for tree diagrams
with quarks, as they do contain $1/N$-suppressed terms in $SU(N)$ gauge theory.}.

In section {\bf 2} we will briefly recall well-known results about decomposition of
full amplitudes into the colour factor $T_n$ and the purely kinematic partial
amplitude $A_n$.
A key point in the approach of \cite{Witten,CSW} is that
only the kinematic amplitude $A_n$ is evaluated directly.
Since $A_n$ does not contain colour factors, it is the same for
tree amplitudes involving quarks and for those with gluinos.

A priori, when comparing
kinematic amplitudes in a non-supersymmetric and in a supersymmetric theory,
we should make sure that both theories have a similar field content.
In particular, when comparing kinematic amplitudes in QCD and in SYM,
(at least initially)
we need to restrict to the SYM theory with vectors, fermions and no scalars.
Scalars are potentially dangerous, since they can propagate in the internal lines
and spoil the agreement between the amplitudes.
Hence,
while the kinematic tree-level amplitudes in massless QCD agree with those
in ${\mathcal  N}=1$ pure SYM, one might ask if the agreement is lost
when comparing QCD amplitudes to amplitudes in
 ${\mathcal  N}=4$ (and ${\mathcal  N}=2$) theories.  Fortunately, this is not the case,
the agreement between multi-particle quark--gluon amplitudes in QCD and
the corresponding gluino--gluon amplitudes
in SYM theories does not depend on ${\mathcal  N}$.
The main point here is that in ${\mathcal  N}=2$ and ${\mathcal  N}=4$ theories, the scalars
$\phi$
couple to gluinos $\Lambda^A$ and $\Lambda^B$ from different
${\mathcal  N}=1$ supermultiplets,
\be
S_{\rm Yukawa} =\ g_{\rm YM}\, \tr \, \Lambda^{-}_A [\phi^{AB}, \Lambda^{-}_B] \, + \,
g_{\rm YM}\, \tr \, \Lambda^{A+} [\overline{\phi}_{AB}, \Lambda^{B+}] \ ,
\ee
where $A,B=1,\ldots {\mathcal  N} $, and $\phi^{AB}=-\phi^{BA}$, hence $A\neq B$.
Meanwhile
quarks are identified with gluinos of the same fixed $A$, i.e.
$q \leftrightarrow\Lambda^{A=1\, +}$,
$\overline{q} \leftrightarrow\Lambda^{-}_{A=1}.$
QCD-amplitudes with $m$ quarks, $m$ antiquarks and $l$
gluons in
external lines correspond to SYM-amplitudes with $m$ gluinos
$\Lambda^{1+}$, $m$ anti-gluinos $\Lambda^{-}_1$,
and $l$ gluons. Since all external (anti)-gluinos
are from the same ${\mathcal  N}=1$ supermultiplet,
they cannot produce scalars in the internal lines of tree diagrams. These
diagrams are all the same for all ${\mathcal  N}=0,\ldots,4$. Of course, in
${\mathcal  N}=4$ and ${\mathcal  N}=2$ theories there are other classes of diagrams with
gluinos from different ${\mathcal  N}=1$ supermultiplets, and also with scalars in external lines.
Applications of the scalar graph approach to
these more general classes of tree amplitudes in ${\mathcal  N}=2,4$ SYM
will be discussed in sections {\bf 5} and {\bf 6}.

Following \cite{GK} we conclude that,
if the CSW formalism gives correct results for partial
amplitudes $A_n$ in a supersymmetric theory,
it will also work in a nonsupersymmetric case,
and for a finite number of colours.
Full amplitudes are then determined uniquely from the kinematic part $A_n$,
and the known expressions for $T_n$, given in Eqs.~\eqref{twohalf}, \eqref{cfqq}
below.
This means \cite{GK} that for tree amplitudes questions (1) and (3) are essentially the same,
and we have a positive answer to question (4).

In section {\bf 3} we will concentrate on tree-level non-MHV (NMHV) amplitudes with gluons only.
We will review the CSW formalism \cite{CSW} for calculating these amplitudes and note
a technical subtlety which occurs in the calculation.
There are unphysical singularities which occur in certain diagrams.
They must cancel between individual contributions to physical amplitudes.
These cancellations
were carried out successfully in all known cases \cite{CSW,GK,Kosower}, but
it is more desirable to avoid them altogether \cite{GGK}.
Purely gluonic amplitudes can be related via supersymmetric Ward identities
to amplitudes containing fermions. The latter are
free of
unphysical singularities for generic phase space points
and no further helicity-spinor algebra is required to
convert the results into an immediately usable form. The main result
of section {\bf 3} is equation \eqref{restnew} which expresses purely gluonic
amplitudes in terms of amplitudes with fermions which are free from unphysical singularities.

Calculations of amplitudes involving fermions
were carried out in \cite{GGK} and will be reproduced in later sections ({\bf 7} and {\bf 8}).
But first we need to set up the formalism for the CSW scalar graph method
in presence of fermions (and scalars). A natural way to do it is to consider
gauge theories with extended supersymmetry.

In section {\bf 4} we will write down ${\mathcal  N}=4$ supersymmetry algebra in helicity basis which
can be used for deriving
supersymmetric Ward identities for generic $1\le {\mathcal  N} \le 4$ gauge theories.

In section {\bf 5} we will present the analytic ${\mathcal  N}=4$ supervertex of Nair \cite{Nair}
which incorporates all the component vertices needed for the scalar graph formalism
in generic gauge theories with gauge fields, fermions and scalars.
We will see that, interestingly, all of the allowed vertices are not MHV
in theories with scalars \cite{GGK}.
For example,
$A_n(g^-,\Lambda^{1+},\Lambda^{2+},\Lambda^{3+},\Lambda^{4+} )$ is an analytic, but non-MHV
amplitude in ${\mathcal  N}=4$ theory.
This implies
that the scalar graph approach is not primarily based on MHV
amplitudes.

In section {\bf 6} we will apply the scalar graph formalism for calculating
three simple examples of $\overline{\rm MHV}$ (or, more precisely, antianalytic) amplitudes
which involve fermions and gluons.
In all cases we will reproduce known
results for these antianalytic amplitudes, which implies that at tree level
the scalar graph
method appears to work correctly not only in ${  {\mathcal  N}}=0,1$ theories, but also in full
${  {\mathcal  N}}=2$ and ${  {\mathcal  N}}=4$ SYM. In particular,
the ${  {\mathcal  N}}=4$ result
\eqref{reaan4}
 verifies the fact that the building blocks of the scalar graph method are
indeed the analytic vertices \eqref{listanal},
which can have less than 2 negative helicities, i.e. are not MHV.

General tree-level expressions for
$n$-point amplitudes
with three negative helicities carried by fermions and gluons were derived in \cite{GGK}.
We will reproduce these calculations in sections {\bf 7} and {\bf 8}.
In section {\bf 9} we will show how to calculate tree amplitudes
with vectors, fermions and scalars compactly using the scalar
graph method with the supervertex of section {\bf 5}.

In section {\bf 10} we will briefly review the known
applications of the CSW scalar graph approach for loops and
section {\bf 11} presents our conclusions.

\bigskip

In the original topological string theory
formulation \cite{Witten},
one obtains tree-level amplitudes with $d+1$ negative-helicity gluons
from contributions of D-instantons of degree $d$. These  include
contributions from connected multi-instantons of degree-$d$ and from
$d$ disconnected single instantons (as well as from all intermediate
mixed cases of total degree $d$).
The CSW formalism \cite{CSW}
is based on integrating only over the moduli space of completely
disconnected instantons of degree one, linked by twistor space
propagators. Degree-one D-instantons in string theory correspond to
MHV vertices, and obtaining amplitudes with an arbitrary
number of negative helicities using MHV vertices is dual in twistor space to
integrating over degree-one instantons.

In Refs.~\cite{RSV1,RSV2} Roiban, Spradlin, and Volovich
computed the
integrals in the opposite regime --
over the moduli space of connected D-instantons of degree $d$.
They found that for $\overline{\rm MHV}$
amplitudes, and certain next-to-MHV amplitudes
these integrals correctly reproduce
known expressions for gauge theory amplitudes.
We therefore seem to have
different ways of computing the amplitudes from the topological B
model.
These different prescriptions were reconciled on the string theory side
in \cite{GLMN} by showing that the corresponding integrals over instanton moduli
spaces can be reduced to an integral over the common boundary and are
hence equivalent.
On the field theory side, the equivalence of different prescriptions was
explained
in \cite{BBK} as the freedom to choose different decompositions of any given
NMHV tree diagram into smaller blocks of MHV and NMHV diagrams.
In this paper we use the CSW formalism for computing gauge theory amplitudes.
On the string theory side this corresponds to
choosing the disconnected instantons prescription.

\section{Tree Amplitudes}

We will consider tree-level amplitudes in a generic $SU(N)$ gauge theory
with an arbitrary
finite number of colours.
$SU(N)$ is unbroken and all fields are
taken to be massless, we refer to them generically as gluons, fermions and
scalars.

\subsection{Colour decomposition}

It is well-known that a full n-point amplitude ${  M}_n $
can be represented as a sum of products of colour factors $T_n$
and purely kinematic partial
amplitudes $A_n$,
\be
{  M}_n (\{k_i,h_i,c_i\}) \,=\, \sum_{\sigma} \,
T_n (\{c_{\sigma(i)}\}) \, A_n (\{k_{\sigma(i)},h_{\sigma(i)}\}) \, .
\label{one}
\ee
Here $\{c_i\}$ are colour labels of external legs $i=1 \ldots n$, and
the kinematic variables $\{k_i,h_i\}$ are on-shell external momenta and helicities:
all $k_i^2=0,$ and $h_i=\pm 1$ for gluons, $h_i=\pm {1\over 2}$ for fermions, and
$h_i=0$ for scalars.
The sum in \eqref{one} is over appropriate simultaneous permutations $\sigma$ of
colour labels $\{c_{\sigma(i)}\}$ and kinematic variables
$\{k_{\sigma(i)},h_{\sigma(i)}\}$.
The colour factors $T_n$ are easy to determine,
and the non-trivial information about the full amplitude
${  M}_n $ is contained in the purely kinematic part $A_n$. If the
partial amplitudes $A_n(\{k_i,h_i\})$ are known for all permutations $\sigma$ of
the kinematic variables, the full amplitude ${  M}_n $ can be determined from
\eqref{one}.

We first consider tree amplitudes with fields in the adjoint representation only
(e.g. gluons, gluinos and no quarks).
The colour variables $\{c_i\}$ correspond to the
adjoint representation indices, $\{c_i\}=\{a_i\},$ and the colour factor $T_n$
is a single trace of generators,
\be
{  M}_n^{\rm tree} (\{k_i,h_i,a_i\})\, =\, \sum_{\sigma}
\tr({\rm T}^{a_{\sigma(1)}} \ldots {\rm T}^{a_{\sigma(n)}})\
A_n^{\rm tree} (k_{\sigma(1)},h_{\sigma(1)}, \ldots, k_{\sigma(n)},h_{\sigma(n)}) \, .
\label{two}
\ee
Here the sum is over $(n-1)!$ noncyclic inequivalent permutations of $n$
external particles. The single-trace structure in \eqref{two},
\be
\label{twohalf}
T_n\, = \, \tr({\rm T}^{a_{1}} \ldots {\rm T}^{a_{n}}) \, ,
\ee
implies
that all tree level amplitudes
of particles transforming in the adjoint
representation of $SU(N)$ are planar. This is not the case
neither for loop amplitudes, nor for tree amplitudes involving fundamental quarks.

Fields in the fundamental representation
couple to the trace $U(1)$ factor of the $U(N)$ gauge group.
In passing to the $SU(N)$ case this introduces power-suppressed $1/N^p$
terms. However, there is a remarkable
simplification for tree diagrams involving fundamental quarks: the factorisation property
\eqref{one} still holds. More precisely, for a fixed colour ordering $\sigma$, the amplitude
with $m$ quark-antiquark pairs and $l$ gluons
is still a perfect product,
\be
T_{l+2m} (\{c_{\sigma(i)}\}) \ A_{l+2m} (\{k_{\sigma(i)},h_{\sigma(i)}\}) \, ,
\label{three}
\ee
and all $1/N^p$ corrections
to the amplitude are contained in the first term. For tree amplitudes
the exact colour factor in
\eqref{three} is \cite{MP}
\be
T_{l+2m} \, = \,
{(-1)^p\over N^p} ({\rm T}^{a_1} \ldots {\rm T}^{a_{l_1}})_{i_1 \alpha_1}
({\rm T}^{a_{l_1+1}} \ldots {\rm T}^{a_{l_2}})_{i_2 \alpha_2} \ldots
({\rm T}^{a_{l_{m-1}+1}} \ldots {\rm T}^{a_{l}})_{i_m \alpha_m} \, .
\label{cfqq}
\ee
Here $l_1, \ldots , l_{m}$ correspond to an arbitrary partition of an arbitrary
permutation of the $l$ gluon indices; $i_1, \ldots i_m$ are
colour indices of quarks, and $\alpha_1, \ldots \alpha_m$ -- of the antiquarks.
In perturbation theory each external quark is connected by a fermion line to an external
antiquark (all particles are counted as incoming). When quark $i_k$ is connected by a fermion
line to antiquark $\alpha_k$, we set $\alpha_k=\bar{i_k}$. Thus, the set of
$\alpha_1, \ldots \alpha_m$ is a permutation of the set
$\bar{i_1}, \ldots \bar{i_m}$. Finally,
the power $p$ is equal to the number of times $\alpha_k=\bar{i_k}$ minus 1.
When there is only one quark-antiquark pair, m=1 and p=0. For a general $m$,
the power $p$ in \eqref{cfqq} varies from $0$ to $m-1$.

The kinematic amplitudes $A_{l+2m}$ in \eqref{three} have the colour information stripped off
and hence do not distinguish between fundamental quarks and adjoint gluinos.
Thus,
\be
A_{l+2m}(q,\ldots,\bar{q},\ldots, g^+,\ldots, g^-)\, = \,
A_{l+2m}(\Lambda^+,\ldots,\Lambda^-,\ldots, g^+,\ldots, g^-)
\label{four}
\ee
where $q$, $\bar{q}$, $g^{\pm}$, $\Lambda^{\pm}$ denote
quarks, antiquarks, gluons and gluinos of $\pm$ helicity.

In following sections we will use
the scalar graph formalism of \cite{CSW} to evaluate the kinematic
amplitudes $A_n$ in \eqref{four}.
Full amplitudes can then be determined uniquely from the kinematic part $A_n$,
and the known expressions for $T_n$ in \eqref{twohalf} and \eqref{cfqq}
by summing over the inequivalent colour orderings in \eqref{one}.

{}From now on we concentrate on the purely kinematic part of the amplitude,
$A_n$.

\subsection{Amplitudes in the spinor helicity formalism}

We will first consider theories with ${  {\mathcal  N}} \le 1$ supersymmetry.
Gauge theories with extended supersymmetry have a more intricate behaviour
of their amplitudes in the helicity basis
and their study will be postponed until section {\bf 4}.
Theories with ${  {\mathcal  N}}=4$ (or ${  {\mathcal  N}}=2$) supersymmetry have ${  {\mathcal  N}}$ different species of gluinos
and 6 (or 4) scalar fields. This leads to a large number of elementary MHV-like
vertices in the scalar graph formalism. This proliferation of elementary vertices
asks for a super-graph generalization of the CSW scalar graph method, which
will be outlined in section {\bf 5} following Ref.~\cite{GGK}.

Here we will concentrate on tree level partial amplitudes $A_n=A_{l+2m}$
with $l$ gluons and $2m$ fermions in the helicity basis, and all
external lines are defined to be
incoming.

In ${  {\mathcal  N}} \le 1$ theory a fermion of helicity $+{1\over 2}$ is
always connected
by a fermion propagator to a helicity $-{1\over 2}$
fermion hence the number
of fermions $2m$ is always even. This statement is correct
only in theories without scalar fields. In the ${  {\mathcal  N}}=4$ theory,
a pair of positive helicity fermions, $\Lambda^{1+}$, $\Lambda^{2+}$,
can be connected to another pair of positive helicity
fermions, $\Lambda^{3+}$, $\Lambda^{4+}$, by a scalar propagator.

In ${  {\mathcal  N}} \le 1$ theory
a tree amplitude $A_n$ with less than two opposite helicities
vanishes\footnote{In the ${  {\mathcal  N}}=1$ theory this is also correct
to all orders in the loop expansion and non-perturbatively.}
identically \cite{Grisaru}.
First nonvanishing amplitudes contain $n-2$ particles
with helicities of the same sign
\cite{PT,BG}
and are called maximal helicity violating
(MHV) amplitudes.

In the spinor helicity formalism
\cite{Berends,PT,BG} an on-shell momentum
of a massless particle, $p_\mu p^\mu=0,$ is represented as
\be
p_{a \dot a} \equiv \ p_\mu \sigma^\mu_{a \dot a}
=\ \lambda_a\tilde\lambda_{\dot a} \ ,
\ee
where $\lambda_a$ and $\tilde\lambda_{\dot a}$
are two commuting
spinors of positive and negative chirality.
Spinor inner products are defined
by\footnote{Our conventions for spinor helicities follow
\cite{Witten,CSW} and are the same as in \cite{GK,GGK}.}
\be
\langle \lambda,\lambda'\rangle = \ \epsilon_{ab}\lambda^a\lambda'{}^b
 \ , \qquad
[\tilde\lambda,\tilde\lambda'] =\ \epsilon_{\dot a\dot b}
\tilde\lambda^{\dot a}\tilde\lambda'{}^{\dot b} \ ,
\ee
and a scalar product of two null vectors,
$p_{a\dot a}=\lambda_a \tilde\lambda_{\dot a}$ and
$q_{a\dot a}=\lambda'_a\tilde\lambda'_{\dot a}$, becomes
\be \label{scprod}
p_\mu q^\mu =\ {1\over 2}
\langle\lambda,\lambda'\rangle[\tilde\lambda,\tilde\lambda'] \ .
\ee

An MHV amplitude $A_n=A_{l+2m}$
with $l$ gluons and $2m$ fermions in ${  {\mathcal  N}} \le 1$ theories
exists only for $m=0,1,2$.
This is because it must have precisely $n-2$ particles with positive and
$2$ with negative helicities, and our fermions always come in pairs
with helicities $\pm {1\over 2}$.
Hence, there are
three types of MHV tree amplitudes in ${  {\mathcal  N}} \le 1$ theories:
\be
\label{threecls}
A_n (g_r^-,g_s^-) \ , \quad
A_n (g_t^-,\Lambda_r^-,\Lambda_s^+)\ , \quad
A_n (\Lambda_t^-,\Lambda_s^+,\Lambda_r^-,\Lambda_q^+) \ .
\ee
Suppressing the overall
momentum conservation factor,
\be
 i g_{\rm YM}^{n-2} \, (2\pi)^4 \, \delta^{(4)} (\sum_{i=1}^n \lambda_{i a}
\tilde{\lambda}_{i \dot a} ) \ ,
\ee
the MHV purely gluonic amplitude reads \cite{PT,BG}:
\be
A_n (g_r^-,g_s^-)=\
{\langle\lambda_r,\lambda_s\rangle^4\over
\prod_{i=1}^n\langle\lambda_i, \lambda_{i+1}\rangle }
\equiv \
{\vev{r~s}^4 \over \prod_{i=1}^n \vev{i~i+1}} \ ,
\label{mpng}
\ee
where $\lambda_{n+1} \equiv \lambda_1$.
The MHV amplitude with two external fermions and $n-2$ gluons is
\SP{
\label{ndcls}
&A_n (g_t^-,\Lambda_r^-,\Lambda_s^+)= \
{\vev{t~r}^3\ \vev{t~s} \over \prod_{i=1}^n \vev{i~i+1}} \ , \\
&A_n (g_t^-,\Lambda_s^+,\Lambda_r^-)= \
-\ {\vev{t~r}^3\ \vev{t~s} \over \prod_{i=1}^n \vev{i~i+1}} \ ,
}
where the first expression corresponds to $r<s$ and the second to $s<r$
(and $t$ is arbitrary).
The MHV amplitudes with four fermions and $n-4$ gluons on external lines are
\SP{
\label{ndcls2}
&A_n (\Lambda_t^-,\Lambda_s^+,\Lambda_r^-,\Lambda_q^+)
= \
{\vev{t~r}^3\ \vev{s~q} \over \prod_{i=1}^n \vev{i~i+1}} \ , \\
&
A_n (\Lambda_t^-,\Lambda_r^-,\Lambda_s^+,\Lambda_q^+)
= \
-\ {\vev{t~r}^3\ \vev{s~q} \over \prod_{i=1}^n \vev{i~i+1}}
 \ .
}
The first expression in \eqref{ndcls2}
corresponds to $t<s<r<q,$ the second -- to $t<r<s<q,$
and there are other similar expressions,
obtained by further permutations of fermions, with the overall
sign determined by the ordering.

Expressions \eqref{ndcls}, \eqref{ndcls2} can be derived from supersymmetric
Ward identities \cite{Grisaru,MP,Dixon},
and we will have more to say about this in section {\bf 5}.
The $\overline{\rm MHV}$ amplitude can be obtained, as always,
by exchanging helicities $+\leftrightarrow -$ and
$\vev{i~j} \leftrightarrow [i~j].$

\section{Gluonic NMHV amplitudes and the CSW method}

The formalism of CSW was developed in \cite{CSW} for calculating
purely gluonic amplitudes at tree level. In this approach
all non-MHV $n$-gluon
amplitudes (including $\overline{\rm MHV}$) are expressed
as sums of tree diagrams in an effective scalar perturbation theory.
The vertices in this theory are the MHV amplitudes \eqref{mpng},
continued off-shell as described below, and connected by scalar
propagators $1/q^2$.
It was shown in \cite{GK,GGK} that the same idea continues to work in
theories with fermions and gluons. Scattering amplitudes are
determined from scalar diagrams with three types of MHV vertices,
\eqref{mpng},\eqref{ndcls} and \eqref{ndcls2}, which are connected to each
other with scalar propagators $1/q^2$.
Also, since we have argued above that at tree level,
supersymmetry is irrelevant,
the method applies to supersymmetric and non-supersymmetric theories
\cite{GK}.

When one leg of an MHV vertex is connected by a propagator
to a leg of another MHV vertex, both legs become internal
to the diagram and have to be continued off-shell. Off-shell continuation
is defined as follows \cite{CSW}:
we pick an arbitrary spinor $\xi_{\sst\rm Ref}^{\dot a}$ and define
$\lambda_a$ for any internal line carrying momentum $q_{a\dot a}$
by
\be \label{ofsh}
\lambda_a=q_{a\dot a}\xi_{\sst \rm Ref}^{\dot a}\ .
\ee
External lines in a diagram remain on-shell, and for them
$\lambda$ is defined in the usual way.
For the off-shell lines,
the same $\xi_{\sst \rm Ref}$ is used  in all diagrams
contributing to a given amplitude.

For practical applications the authors of \cite{CSW} have chosen
$\xi_{\sst \rm Ref}^{\dot a}$ in \eqref{ofsh}
to be equal to $\tilde{\lambda}^{\dot a}$
of one of the external legs of negative helicity, e.g. the first one,
\be \label{ofsh2}
\xi_{\sst \rm Ref}^{\dot a}=\ \tilde{\lambda}_1^{\dot a}
\ .
\ee
This corresponds to
identifying the reference spinor with one of the kinematic variables of the
theory. The explicit dependence on the reference spinor $\xi_{\sst \rm Ref}^{\dot a}$
disappears and the
resulting expressions for all scalar diagrams in the CSW approach are the functions
only of the kinematic variables $\lambda_{i\, a}$ and $\tilde{\lambda}^{\dot a}_i.$
This means that the expressions for all individual diagrams automatically
appear to be Lorentz-invariant
(in the sense that they do not depend on an external spinor $\xi_{\sst \rm Ref}^{\dot a}$)
and also gauge-invariant (since the reference spinor corresponds to  the
axial gauge fixing $\xi_{\sst \rm Ref}^\mu A_\mu =0,$ where
$\xi_{\sst \rm Ref}^{\dot a a} = \xi_{\sst \rm Ref}^{\dot{a}}\xi_{\sst \rm Ref}^a$).

There is a price to pay for this invariance of the individual diagrams.
Equations~\eqref{ofsh},\eqref{ofsh2} lead to unphysical
singularities\footnote{Unphysical means that these singularities are not the
standard IR soft and collinear divergences in the amplitudes.}
which occur for the whole of phase space and
which have to be cancelled between the individual diagrams. The result for the
total amplitude is, of course, free of these unphysical singularities, but their cancellation
and the retention of the finite part requires some work, see \cite{CSW}
and section 3.1 of \cite{GK}.

Following \cite{GK,GGK} we  note that
these unphysical singularities are specific to the three-gluon MHV vertices
and, importantly, they do not occur in any of the MHV vertices involving a fermion
field.
To see how these singularities arise in gluon vertices, consider a 3-point MHV vertex,
\EQ{
A_3(g_1^-,g_2^-,g_3^+)=\ \frac{\vev{1~2}^4}{\vev{1~2}\vev{2~3}\vev{3~1}}
=\ \frac{\vev{1~2}^3}{\vev{2~3}\vev{3~1}}
\ . \label{simpl1}}
This vertex exists only when one of the legs is off-shell. Take it to be
the $g_3^+$ leg. Then Eqs.~\eqref{ofsh}, \eqref{ofsh2}, and momentum conservation,
$q=p_1+p_2$,  give
\EQ{
\lambda_{3\,a}=\ (p_1+p_2)_{a\dot{a}}\, \tilde{\lambda}_1^{\dot a}=\
-\lambda_{1\, a}\,[1~1] - \, \lambda_{2\, a}\,[2~1] =\
 - \, \lambda_{2\, a}\,[2~1] \ .
}
This implies that $\vev{2~3}=-\vev{2~2}[2~1]=0,$
and the denominator of \eqref{simpl1} vanishes.
This is precisely the singularity we are after.
If instead of the $g_3^+$ leg, one takes the
$g_2^-$ leg go off-shell, then, $\vev{2~3}=-\vev{3~3}[3~1]=0$ again.

Now consider a three-point MHV vertex involving two fermions and a gluon,
\EQ{
A_3(\Lambda_1^-,g_2^-,\Lambda_3^+)=\ \frac{\vev{2~1}^3\vev{2~3}}{\vev{1~2}\vev{2~3}\vev{3~1}}
=\ -\, \frac{\vev{2~1}^2}{\vev{3~1}}
\ . \label{simpl2}}
Choose the reference spinor to be as before, $ \tilde{\lambda}_1^{\dot a},$ and
take the second or the third leg off-shell. This again makes $\vev{2~3}=0$,
but now the factor of $\vev{2~3}$ is cancelled on the right hand side of \eqref{simpl2}.
Hence, the vertex \eqref{simpl2} is regular, and
there are no unphysical singularities
in the amplitudes involving at least one negative helicity fermion when it's helicity
is chosen to be the reference spinor \cite{GK}.
One concludes that the difficulties with singularities at intermediate stages
of the calculation occur only in purely gluonic amplitudes. One way
to avoid these intermediate singularities is to choose an off-shell continuation
different from the CSW prescription \eqref{ofsh},\eqref{ofsh2}.

Recently,
Kosower \cite{Kosower} used an off-shell continuation by projection of the
off-shell momentum with respect to an on-shell reference momentum
$q_{\sst \rm Ref}^\mu,$ to derive, for the first time,
an expression for a general NMHV amplitude
with three negative helicity gluons. The amplitude in \cite{Kosower} was
from the start free of unphysical divergences, however it required
a certain amount of spinor algebra to bring it into the form
independent of the reference momentum.

In \cite{GGK} we proposed another simple method for finding all purely gluonic NMHV
amplitudes. Using ${\mathcal  N}=1$ supersymmetric Ward identities one can relate
purely gluonic amplitudes to a linear combination of amplitudes with
one fermion--antifermion pair. As explained above,
the latter are free of singularities and are
manifestly Lorentz-invariant.
These fermionic amplitudes will be calculated in section {\bf 7}
using the CSW scalar graph approach with fermions \cite{GK}
and following \cite{GGK}.

To derive supersymmetric Ward identities
\cite{Grisaru} we use the fact that,
supercharges $Q$ annihilate the vacuum,
and consider the following equation,
\EQ{
\langle [Q\, , \,
\Lambda^+_k \ldots g_{r_1}^- \ldots g_{r_2}^- \ldots g_{r_3}^- \ldots ]\rangle
\ = \ 0
\ , }
where dots indicate positive helicity gluons. In order to make
anticommuting spinor $Q$ to be a singlet entering a commutative
(rather than anticommutative) algebra
with all the fields we contract it with a commuting spinor $\eta$ and
multiply it by a Grassmann number $\theta$. This defines a
commuting singlet operator $Q(\eta).$
Following \cite{Dixon} we can write down the following susy algebra relations,
\SP{\label{susyward}
[Q(\eta) \, , \, \Lambda^{+}(k)] \ = \ - \theta \vev{\eta~k}\,g^+ (k) \ , \quad
[Q(\eta) \, , \, \Lambda^{-}(k)] \ = \ + \theta [\eta~k]\,g^- (k) \ , \\
[Q(\eta) \, , \, g^{-}(k)] \ = \ + \theta \vev{\eta~k}\,\Lambda^- (k) \ , \quad
[Q(\eta) \, , \, g^{+}(k)] \ = \ - \theta [\eta~k]\,\Lambda^+ (k) \ .
}
In what follows, the anticommuting parameter
$\theta$ will cancel from the relevant expressions for the amplitudes.
The arbitrary spinors  $\eta_a,$ $\eta_{\dot a},$  will be fixed below.
It then follows from \eqref{susyward} that
\SP{\label{restggg}
{\vev{\eta~k}}\, &A_n (g_{r_1}^- , g_{r_2}^- , g_{r_3}^-) = \
{\vev{\eta~r_1}}\,
A_n (\Lambda_{k}^+, \Lambda_{r_1}^-, g_{r_2}^-, g_{r_3}^-)\\
&+ \,
{\vev{\eta~r_2}}\,
A_n (\Lambda_{k}^+, g_{r_1}^-, \Lambda_{r_2}^-, g_{r_3}^-)
+ \,
{\vev{\eta~r_3}}\,
A_n (\Lambda_{k}^+, g_{r_1}^-, g_{r_2}^-, \Lambda_{r_3}^- ) \ .
}
After choosing $\eta$ to be one of the three $r_j$ we find from
\eqref{restggg} that the purely gluonic amplitude with three negative helicities
is given by a sum of two fermion-antifermion-gluon-gluon amplitudes.
Note that in the expressions above and in what follows, in $n$-point amplitudes
we show only the relevant particles, and suppress all the positive
helicity gluons $g^+$.

Remarkably, this approach works for any number of negative helicities,
and the NMHV amplitude with $h$ negative gluons is expressed via a simple
linear combination of $h-1$ NMHV amplitudes with one fermion-antifermion pair.

In  sections {\bf 7} and {\bf 8} we will evaluate NMHV amplitudes with fermions.
In particular,
in section {\bf 7} we will calculate the following three amplitudes,
\EQ{\label{nprocesses}
A_n(\L_{m_1}^-,g_{m_2}^-,g_{m_3}^-,\L_k^+)\ , \quad
A_n(\L_{m_1}^-,g_{m_2}^-,\L_k^+,g_{m_3}^-)\ , \quad
A_n(\L_{m_1}^-,\L_k^+,g_{m_2}^-,g_{m_3}^-) \ .
}
In terms of these, the purely gluonic amplitude of \eqref{restggg}
reads
\SP{\label{restnew}
A_n (g_{r_1}^- , g_{r_2}^- , g_{r_3}^-) = \
-\frac{\vev{\eta~r_1}}{\vev{\eta~k}}\
A_n(\L_{m_1}^-,g_{m_2}^-,g_{m_3}^-,\L_k^+)|_{{m_1}=r_1, m_2=r_2, m_3=r_3} \\
- \,
\frac{\vev{\eta~r_2}}{\vev{\eta~k}}\
A_n(\L_{m_1}^-,g_{m_2}^-,\L_k^+,g_{m_3}^-)|_{{m_1}=r_2, m_2=r_3, m_3=r_1} \\
- \,\frac{\vev{\eta~r_3}}{\vev{\eta~k}}\
A_n(\L_{m_1}^-,\L_k^+,g_{m_2}^-,g_{m_3}^-)|_{{m_1}=r_3, m_2=r_1, m_3=r_2}
 \ ,
}
 and $\eta$ can be chosen to be one of the three $m_j$ to further
 simplify this formula.

\section{${\mathcal  N}=4$ Supersymmetry Algebra in Helicity Formalism}

The ${\mathcal  N}=1$ susy algebra relations \eqref{susyward} can be generalised to
${\mathcal  N}\ge 1$ theories. The ${\mathcal  N}=4$ susy relations
read:
\AL{\label{swa}
&[Q^A(\eta) \, , \, g^{+}(k)] \ = \ - \theta_A [\eta~k]\,\Lambda^{+\,A} (k) \ , \\
\label{swb}
&[Q^A(\eta) \, , \, \Lambda^{+\,B}(k)] \ = \
- \delta^{AB}\,\theta_A \vev{\eta~k}\,g^+ (k)\,-\,\theta_A [\eta~k]\,\phi^{AB}
 \ , \\
 \label{swc}
&[Q^A(\eta) \, , \, \overline{\phi}_{AB}(k)] \ = \ - \theta_A [\eta~k]\,\Lambda^{-}_{B} (k) \ , \\
\label{swd}
&[Q_A(\eta) \, , \, {\phi}^{AB}(k)] \ = \ \theta_A \vev{\eta~k}\,\Lambda^{+\,B} (k)   \ , \\
\label{swe}
&[Q_A(\eta) \, , \, \Lambda^{-}_B (k)] \ = \  \delta_{AB}\,\theta_A [\eta~k]\,g^- (k)
\,+\, \theta_A \vev{\eta~k} \,\overline{\phi}_{AB}(k)    \ , \\
\label{swf}
&[Q_A(\eta) \, , \, g^{-}(k)] \ = \  \theta_A \vev{\eta~k}\,\Lambda^{-}_A (k) \ .
}
Our conventions are the same as in \eqref{susyward}, and it is understood that
$Q_A=Q^A$ and there is no summation over $A$ in \eqref{swc}, \eqref{swd}.
The conjugate scalar field is defined in the standard way,
\be
\overline{\phi}_{AB} \ = \ \hf\, \epsilon_{ABCD}\, \phi^{CD}\ =\
(\phi^{AB})^{\dagger} \ .
\ee
Relations \eqref{swa}-\eqref{swf} can be used in order to derive ${\mathcal  N}=2$
and ${\mathcal  N}=4$ susy Ward identities which relate different classes of
amplitudes in gauge theories
with extended supersymmetry.

\section{The Analytic Supervertex in ${\mathcal  N}=4$ SYM}

So far we have encountered three types of MHV amplitudes
\eqref{mpng}, \eqref{ndcls} and \eqref{ndcls2}. The key feature
which distinguishes these amplitudes is the fact that they
depend only on $\vev{\lambda_i~\lambda_j}$
spinor products,
and not on $[\tilde\lambda_i~ \tilde\lambda_i].$ We will call
such amplitudes analytic.

All analytic amplitudes in generic $0\le {  {\mathcal  N}} \le 4$
gauge theories can be combined into a single ${  {\mathcal  N}}=4$ supersymmetric
expression
of Nair \cite{Nair},
\be
A_n^{{  {\mathcal  N}}=4} =\
\delta^{(8)} \left(\sum_{i=1}^n \lambda_{i a}
\eta^A_i \right)\
{1 \over \prod_{i=1}^n \vev{i~i+1}} \ .
\label{nair}
\ee
Here $\eta^A_i$ are anticommuting variables and $A=1,2,3,4$.
The Grassmann-valued delta function is defined in the usual way,
\EQ{\label{delta8}
\delta^{(8)} \left(\sum_{i=1}^n \lambda_{i a}
\eta^A_i \right) \equiv \ \prod_{A=1}^4 \, \hf
\left(\sum_{i=1}^n \lambda_{i }^a \eta^A_i \right)
\left(\sum_{i=1}^n \lambda_{i a} \eta^A_i \right) \ ,
}
Taylor expanding \eqref{nair} in powers of $\eta_i$, one can identify
each term in the expansion with a particular tree-level analytic amplitude
in the ${  {\mathcal  N}}=4$ theory.  $(\eta_i)^k$ for $k=0,\ldots,4$ is interpreted as
the $i^{\rm th}$ particle with helicity $h_i=1-{k\over 2}$.
This implies that helicities take values,
$\{1,{1\over 2},0,-{1\over 2},-1\},$ which precisely correspond to those of
the ${  {\mathcal  N}}=4$ supermultiplet,
$\{g^-,\lambda^{-}_A,\phi^{AB},\Lambda^{A+},g^+\}.$

It is straightforward to write down a general rule \cite{GK}
for associating a power of
$\eta$ with all component fields in ${  {\mathcal  N}}=4$,
\EQ{ \label{nrules}
g^{-}_i \sim\, \eta_i^1 \eta_i^2 \eta_i^3 \eta_i^4 \ , \
\Lambda^{-}_{1} \sim\,  -\,\eta_i^2 \eta_i^3 \eta_i^4 \ , \
\phi^{AB}_i \sim\,  \eta_i^A \eta_i^B \ , \
\Lambda^{A+}_i \sim\,  \eta_i^A \ , \
g^{+}_i\sim\,  1  \ ,
}
with expressions for the remaining $\Lambda^{-}_{A}$ with
$A=2,3,4$ written in the same manner as the expression for
$\Lambda^{-}_{1}$ in \eqref{nrules}.
The first MHV amplitude \eqref{mpng} is derived from \eqref{nair}
by using the dictionary \eqref{nrules}
and by selecting
the $(\eta_r)^4 \ (\eta_s)^4$ term in \eqref{nair}.
The second amplitude \eqref{ndcls}
follows from the $(\eta_t)^4 (\eta_r)^3 (\eta_s)^1$ term in \eqref{nair};
and the third amplitude
\eqref{ndcls2} is an
$(\eta_r)^3 (\eta_s)^1 (\eta_t)^3 (\eta_q)^1$ term.

There is a large number of such component amplitudes for an
extended susy Yang-Mills, and what is remarkable, not all of these
amplitudes are MHV. The analytic amplitudes of the ${  {\mathcal  N}}=4$ SYM
obtained from \eqref{nair}, \eqref{nrules} are \cite{GGK}:
\SP{\label{listanal}
&A_n(g^-, g^-) \ , \quad
A_n(g^-,\Lambda_A^-,\Lambda^{A+}) \ , \quad
A_n(\Lambda_A^-,\Lambda_B^-,\Lambda^{A+},\Lambda^{B+}) \ , \\
&A_n(g^-,\Lambda^{1+},\Lambda^{2+},\Lambda^{3+},\Lambda^{4+} ) \ , \quad
A_n(\Lambda_A^-,\Lambda^{A+},\Lambda^{1+},\Lambda^{2+},\Lambda^{3+},\Lambda^{4+})
 \ ,   \\
&A_n(\Lambda^{1+},\Lambda^{2+},\Lambda^{3+},\Lambda^{4+},
\Lambda^{1+},\Lambda^{2+},\Lambda^{3+},\Lambda^{4+} ) \ , \\
&A_n(\overline{\phi}_{AB},\Lambda^{A+},\Lambda^{B+},\Lambda^{1+},\Lambda^{2+},\Lambda^{3+},\Lambda^{4+})
 \ ,  \\
&A_n(g^-,\overline{\phi}_{AB}, \phi^{AB}) \ , \quad
A_n(g^-,\overline{\phi}_{AB}, \Lambda^{A+},\Lambda^{B+}) \ , \quad
A_n(\Lambda_A^-,\Lambda_B^-, \phi^{AB}) \ , \\
&A_n(\Lambda_A^-,{\phi}^{AB},\overline{\phi}_{BC},\Lambda^{C+}) \ , \quad
A_n(\Lambda_A^-,\overline{\phi}_{BC},\Lambda^{A+}, \Lambda^{B+},\Lambda^{C+})
 \ , \\
&A_n( \overline{\phi}, {\phi}, \overline{\phi}, {\phi}) \ , \quad
A_n( \overline{\phi}, {\phi}, \overline{\phi},\Lambda^{+}, \Lambda^{+}) \ , \quad
A_n( \overline{\phi},\overline{\phi},\Lambda^{+}, \Lambda^{+},
\Lambda^{+}, \Lambda^{+}) \ ,
 }
where it is understood that
$\overline{\phi}_{AB}=\hf \epsilon_{ABCD}\phi^{CD}.$
In Eqs.~\eqref{listanal} we do not distinguish between the different particle orderings in the
amplitudes. The labels refer to supersymmetry multiplets, $A,B=1,\ldots,4.$
Analytic amplitudes in \eqref{listanal} include the familiar MHV amplitudes,
 \eqref{mpng}, \eqref{ndcls}, \eqref{ndcls2}, as well as more complicated classes
of amplitudes with external gluinos $\Lambda^A,$ $\Lambda^{B\neq A},$ etc, and
with external scalar fields $\phi^{AB}.$

The second, third and fourth lines in \eqref{listanal} are not even MHV
amplitudes, they have less than two negative helicities, and nevertheless,
these amplitudes are non-vanishing  in ${  {\mathcal  N}}=4$ SYM.
The conclusion we draw \cite{GGK} is
that in the scalar graph formalism in ${  {\mathcal  N}} \le 4$ SYM, the
amplitudes are characterised not by a number of negative helicities,
but rather by the total number of $\eta$'s associated to each amplitude
via the rules \eqref{nrules}.

All the analytic amplitudes listed in \eqref{listanal} can be calculated
directly from \eqref{nair}, \eqref{nrules}. There is a simple algorithm
for doing this \cite{GGK}.
\begin{enumerate}
\item{} For each amplitude in \eqref{listanal} substitute the
fields by their $\eta$-expressions \eqref{nrules}. There are precisely
eight $\eta$'s for each analytic amplitude.
\item{} Keeping track of the overall sign, rearrange the anticommuting
$\eta$'s into a product of four pairs:
$({\rm sign})\times
\eta^1_i \eta^1_j \,\eta^2_k \eta^2_l \,\eta^3_m \eta^3_n \,\eta^4_r \eta^4_s.$
\item{} The amplitude is obtained by replacing each pair $\eta^A_i \eta^A_j$
by the spinor product $\vev{i~j}$ and dividing by the usual denominator,
\EQ{\label{analsimp}
A_n = \ ({\rm sign})\times
\frac{\vev{i~j}\vev{k~l}\vev{m~n}\vev{r~s}}{\prod_{l=1}^n\ \vev{l~l+1}}
\ .
}
\end{enumerate}
In this way one can immediately write down expressions for
all component amplitudes in \eqref{listanal}.
It can be checked that these expressions
are inter-related via ${\mathcal  N}=4$ susy Ward identities
which follow from the ${\mathcal  N}=4$ susy algebra in \eqref{swa}-\eqref{swf}.

The vertices of the scalar graph method are the analytic vertices
\eqref{listanal} which are all of degree-8 in $\eta$ and are not
necessarily MHV. These are component vertices of a single analytic
supervertex\footnote{The list of component vertices \eqref{listanal} is obtained
by writing down all partitions of 8 into groups of 4, 3, 2 and 1.
For example, $A_n(g^-,\overline{\phi}_{AB}, \Lambda^{A+},\Lambda^{B+})$
follows from $8=4+2+1+1.$}
\eqref{nair}.
The analytic amplitudes of degree-8 are the elementary
blocks of the scalar graph approach. The next-to-minimal case are
the amplitudes of degree-12 in $\eta$, and they are obtained by connecting
two analytic vertices of \cite{Nair} with a scalar propagator $1/q^2.$
Each analytic vertex
contributes 8 $\eta$'s and a propagator removes 4. Scalar diagrams with
three degree-8 vertices give the degree-12 amplitude, etc.
In general,
all $n$-point amplitudes are characterised by a degree $8,12, 16, \ldots, (4n-8)$
which are obtained from scalar diagrams with $1,2,3, \ldots$ analytic
vertices.\footnote{In practice, one needs to know only the first half of these amplitudes,
since degree-$(4n-8)$ amplitudes are anti-analytic (also known as
googly)  and they are simply given by degree-$8^*$ amplitudes,
similarly degree-$(4n-12)$ are given by degree-$12^*$, etc.}
In section {\bf 9} we will derive a simple
expression for the first iteration of the degree-8 vertex. This iterative
process can be continued straightforwardly to higher orders.

\section{Calculating Simple Antianalytic Amplitudes}

To show the simplicity of the scalar graph method at tree level and to test its results,
in this section
we will calculate some simple antianalytic amplitudes
of $\eta$-degree-12. More complicated general cases are discussed in sections
{\bf 7} -- {\bf 9}.

We work in
${  {\mathcal  N}}=1,$ ${  {\mathcal  N}}=2$ and ${  {\mathcal  N}}=4$ SYM theories, and study
\AL{ \label{mhvb1}
A^{{  {\mathcal  N}}=1}_5(g_1^-,\Lambda_{(1)\,2}^-,\Lambda_{(1)\,3}^-,
\Lambda_{4}^{(1)+},\Lambda_5^{(1)+}) \ , \\
\label{mhvb2}
A^{{  {\mathcal  N}}=2}_5(\Lambda_{(1)\,1}^-,\Lambda_{(2)\,2}^-,g_3^-,
\Lambda_{4}^{(1)+},\Lambda_5^{(2)+}) \ , \\
\label{mhvb4}
A^{{  {\mathcal  N}}=4}_5(\Lambda_{(1)\,1}^-,\Lambda_{(2)\,2}^-,
\Lambda_{(3)\,3}^-,\Lambda_{(4)\,4}^-,g^+_5) \ ,
}
using the scalar graph method with analytic vertices. The labels
${  {\mathcal  N}}=1,2,4$ on the three amplitudes above corresponds to the {\it minimal}
number of supersymmetries for the given amplitude.
In this section
the ${  {\mathcal  N}}$-supersymmetry labels $A,B$ are shown as $(A)$ and $(B)$.

In all cases we will reproduce known
results for these antianalytic amplitudes, which implies that at tree level
the scalar graph
method appears to work correctly not only in ${  {\mathcal  N}}=0,1$ theories, but also in full
${  {\mathcal  N}}=2$ and ${  {\mathcal  N}}=4$ SYM. This answers the question (2) in the introduction.

Furthermore, the ${  {\mathcal  N}}=4$ result
\eqref{reaan4}
for the amplitude \eqref{mhvb4}
will verify the fact that the building blocks of the scalar graph method are
indeed the analytic vertices \eqref{listanal},
which can have less than 2 negative helicities, i.e. are not MHV.

We will be using the of-shell prescription
$\xi_{\sst \rm Ref}^{\dot a}=\ \tilde{\lambda}_2^{\dot a}$
as in the section {\bf 3}.
Since in our amplitudes, the reference spinor $\tilde\lambda_2^{\dot a}$ always
corresponds to a gluino $\Lambda-$,
rather than a gluon $g^-$, there will be no singularities
in our formulae at any stage of the calculation.

\subsection{Antianalytic  ${  {\mathcal  N}}=1$ amplitude}

There are three diagrams contributing to the first amplitude, Eq.~\eqref{mhvb1}.
The first one is a gluon exchange between two
2-fermion MHV-vertices. This diagram has a schematic form,
\be \label{beone}
A_4(g_1^-,\Lambda_2^-,\underline{g_I^+},\Lambda_5^+) \ {1\over q_I^2} \
A_3(\Lambda_3^-,\Lambda_4^+,\underline{g_{-I}^-}) \ .
\ee
Here $\underline{g_I^+}$ and $\underline{g_{-I}^-}$ are off-shell (internal)
gluons which are Wick-contracted via a scalar
propagator, and $I=(3,4)$, which means,
$\lambda_I=(p_3+p_4)\cdot \tilde\lambda_2$.

The second and the third diagrams involve a fermion exchange
between a 2-fermion and a 4-fermion MHV vertices. They are given, respectively
by
\be \label{feone}
A(\Lambda_2^-,\Lambda_3^-,\Lambda_4^+,\underline{\Lambda_{-I}^+})
\ {1\over q_I^2} \
A(\Lambda_5^+,g_1^-,\underline{\Lambda_I^-}) \ ,
\ee
with $I=(2,4),$ and
\be \label{fetwo}
A(g_1^-,\Lambda_2^-,\underline{\Lambda_I^+})\ {1\over q_I^2} \
A(\Lambda_3^-,\Lambda_4^+,\Lambda_5^+,\underline{\Lambda_{-I}^-}) \ ,
\ee
with $I=(3,5).$
Both expressions, \eqref{feone} and \eqref{fetwo}, are written in the form
which is in agreement with the ordering prescription of \cite{GK} for internal
fermions, ket$^+$ ket$^-$.
All three contributions are straightforward to evaluate using the
relevant expressions for the component analytic vertices. These expressions follow
from the algorithm \eqref{analsimp}.

1. The first contribution, Eq.~\eqref{beone}, is
\SP{ \label{one1}
&{-\vev{1~2}^2 \over (\vev{2~3}[2~3]+\vev{2~4}[2~4])\vev{5~1}[2~1]}  \, \cdot \,
{1 \over \vev{3~4}[3~4]}  \, \cdot \,
\vev{4~3}[2~4]^2 \\
&=\
{[2~4]^2\vev{1~2}^2 \over [3~4] (\vev{2~3}[2~3]+\vev{2~4}[2~4])\vev{5~1}[2~1]} \ .
}

2. The second diagram, Eq.~\eqref{feone}, gives
\SP{ \label{three3}
&{-\vev{2~3}^2 \over (\vev{2~3}[2~3]+\vev{2~4}[2~4])\vev{3~4}} \, \cdot \,
{1 \over \vev{5~1}[5~1]} \, \cdot \,
{\vev{5~1}[2~5]^2 \over [2~1]} \\
&=\ {-[2~5]^2\vev{2~3}^2 \over [2~1] [5~1] (\vev{2~3}[2~3]+\vev{2~4}[2~4])\vev{3~4}} \ .
}

3. The third contribution, Eq.~\eqref{fetwo}, is
\be
 \label{four4}
 {\vev{2~1}\over [2~1]}  \, \cdot \,
{1 \over \vev{1~2}[1~2]}
 \, \cdot \,
{\vev{3~1}^2[2~1] \over \vev{3~4}\vev{5~1}} \ = \
{\vev{3~1}^2 \over [2~1] \vev{3~4}\vev{5~1}} \ .
\ee

Now, we need to add up the three contributions.
We first combine the expressions in \eqref{one1} and
\eqref{three3} into
\be
{[4~5]^2 \over [2~1][3~4][5~1]} -
{\vev{3~1}^2 \over [2~1] \vev{3~4}\vev{5~1}}
\ee
using momentum conservation identities, and the fact that
$\vev{2~3}[2~3]+\vev{2~4}[2~4]=-\vev{3~4}[3~4]+\vev{5~1}[5~1].$
Then, adding the remaining contribution \eqref{four4} we
obtain the final result for the amplitude,
\be
A^{{  {\mathcal  N}}=1}_5(g_1^-,\Lambda_{(1)\,2}^-,\Lambda_{(1)\,3}^-,
\Lambda_{4}^{(1)+},\Lambda_5^{(1)+})
\, = \,
{-[4~5]^3[2~3] \over [1~2][2~3][3~4][4~5][5~1]} \ .
\ee
which is precisely the right answer for the antianalytic
5-point `mostly minus' diagram.
This can be easily verified by
taking a complex conjugation (parity transform)
of the corresponding analytic expression.

\subsection{Antianalytic  ${  {\mathcal  N}}=2$ amplitude}

There are three contributions to the amplitude
\eqref{mhvb2}
The first contribution is a scalar exchange between two
analytic vertices,
\be \label{be2one}
A_3(\Lambda_{(1)\,1}^-,\Lambda_{(2)\,2}^-,\underline{\phi^{(12)}_{-I}}) \ {1\over q_I^2} \
A_4(g_3^-,\Lambda_{4}^{(1)+},\Lambda_5^{(2)+},\underline{\phi_{(12)\,I}}) \ .
\ee
Here $\underline{\phi^{(12)}_{-I}}$ and
$\underline{\phi_{(12)\,I}}\equiv \underline{\phi^{(34)}_{I}}$ are off-shell (internal)
scalars which are Wick-contracted.
The external index $I=(1,2),$ which implies  $\lambda_I=(p_1+p_2)\cdot \tilde\lambda_2=
p_1\cdot \tilde\lambda_2$.
The second contribution to \eqref{mhvb2} is a fermion exchange,
\be \label{fe2one}
A_3(g_3^-,\Lambda_{4}^{(1)+},\underline{\Lambda_{(1)\,-I}^-}) \ {1\over q_I^2} \
A_4(\Lambda_5^{(2)+},\Lambda_{(1)\,1}^-,\Lambda_{(2)\,2}^-,
\underline{\Lambda^{(1)+}_{I}})
 \ ,
\ee
with external index $I=(3,4),$ that is, $\lambda_I=(p_3+p_4)\cdot \tilde\lambda_2$.

The final third contribution is again a fermion exchange,
\be \label{fe2two}
A_3(\Lambda_{(2)\,2}^-,g_3^-,\underline{\Lambda^{(2)+}_{-I}})
\ {1\over q_I^2} \
A_4(\Lambda_{4}^{(1)+},\Lambda_5^{(2)+},\Lambda_{(1)\,1}^-,
\underline{\Lambda_{(2)\,I}^-} ) \ ,
\ee
with $I=(2,3),$ and $\lambda_I=(p_2+p_3)\cdot \tilde\lambda_2=
p_3\cdot \tilde\lambda_2$.
As before, all three contributions are straightforward to evaluate using
 the rules \eqref{analsimp}.

1. The first contribution, Eq.~\eqref{be2one}, is
\EQ{ \label{2one1}
\vev{1~2} \, \cdot \,
{1 \over \vev{1~2}[1~2]}  \, \cdot \,
\frac{-\vev{3~5}\vev{3~1}[2~1]}{\vev{4~5}\vev{5~1}[2~1]} \,
=\
\frac{\vev{3~5}\vev{3~1}}{\vev{4~5}\vev{5~1}} \frac{1}{[2~1]}
 \ .
}

2. The second contribution \eqref{fe2one} gives
\EQ{ \label{2two1}
\frac{\vev{3~4}^2 [2~4]^2}{\vev{4~3}[2~3]}
 \, \cdot \,
{1 \over \vev{3~4}[3~4]}  \, \cdot \,
\frac{-\vev{1~2}}{\vev{5~1}[2~1]}
\,=\
- \ \frac{\vev{1~2}}{\vev{5~1}} \frac{[2~4]^2}{[1~2][2~3][3~4]}
 \ .
}

3. The third  contribution \eqref{fe2two}, is
\EQ{ \label{2three3}
-\, \frac{\vev{2~3}^2}{\vev{2~3}[3~2]}\, \cdot \,
{1 \over \vev{2~3}[2~3]}  \, \cdot \,
\frac{\vev{1~3}[2~3]}{\vev{4~5}} \,=\
 \frac{\vev{1~3}}{\vev{4~5}} \frac{1}{[2~3]}
\ .
}

Now, we  add up the three contributions in
Eqs.~\eqref{2one1}, \eqref{2two1}, \eqref{2three3}
and using the momentum conservation identities obtain
\be
A^{{  {\mathcal  N}}=2}_5(\Lambda_{(1)\,1}^-,\Lambda_{(2)\,2}^-,g_3^-,
\Lambda_{4}^{(1)+},\Lambda_5^{(2)+})= \
\frac{[2~4][4~5]}{[1~2][2~3][3~4]} \ ,
\ee
which is the correct result for the antianalytic amplitude.

\subsection{Antianalytic  ${  {\mathcal  N}}=4$ amplitude}

The amplitude
$A^{{  {\mathcal  N}}=4}_5(\Lambda_{(1)\,1}^-,\Lambda_{(2)\,2}^-,
\Lambda_{(3)\,3}^-,\Lambda_{(4)\,4}^-,g^+_5)$
receives contributions only from diagrams with a scalar exchange. There are
three such diagrams.

The first one is
\be \label{be3one}
A_4(g^+_5,\Lambda_{(1)\,1}^-,\Lambda_{(2)\,2}^-,
\underline{\phi^{(12)}_{-I}}) \ {1\over q_I^2} \
A_3(\Lambda_{(3)\,3}^-,\Lambda_{(4)\,4}^-,
\underline{\phi^{(34)}_{I}}) \ .
\ee
Here $\underline{\phi^{(12)}_{-I}}$ and
$\underline{\phi^{(34)}_{I}}$ are off-shell (internal)
scalars which are Wick-contracted and
$\lambda_I=(p_1+p_2+p_5)\cdot \tilde\lambda_2=
(p_1+p_5)\cdot \tilde\lambda_2$.

The second contribution to \eqref{mhvb4} is
\be \label{fe3one}
A_3(\Lambda_{(1)\,1}^-,\Lambda_{(2)\,2}^-
,\underline{\phi^{(12)}_{-I}}) \ {1\over q_I^2} \
A_4(\Lambda_{(3)\,3}^-,\Lambda_{(4)\,4}^-,g^+_5,
\underline{\phi^{(34)}_{I}}) \ .
\ee
with external index $I=(1,2),$ that is, $\lambda_I=(p_1+p_2)\cdot \tilde\lambda_2
=p_1\cdot \tilde\lambda_2$.

The third diagram gives,
\be \label{fe3two}
A_3(\Lambda_{(2)\,2}^-,\Lambda_{(3)\,3}^-
,\underline{\phi^{(23)}_{-I}}) \ {1\over q_I^2} \
A_4(\Lambda_{(4)\,4}^-,g^+_5,\Lambda_{(1)\,1}^-,
\underline{\phi^{(14)}_{I}}) \ ,
\ee
with $I=(2,3),$ and $\lambda_I=(p_2+p_3)\cdot \tilde\lambda_2=
p_3\cdot \tilde\lambda_2$.

1. The first contribution, Eq.~\eqref{be3one}, is
\EQ{ \label{3one1}
\frac{\vev{1~5}[2~5]\vev{1~2}}{\vev{5~1}[2~1]\vev{5~1}}
\, \cdot \,
{1 \over \vev{3~4}[3~4]}  \, \cdot \,
\vev{3~4} \,
=\
 \frac{\vev{1~2}}{\vev{1~5}}
 \frac{[2~5]}{[2~1][3~4]}
 \ .
}

2. The second contribution \eqref{fe3one} gives
\EQ{ \label{3two1}
\vev{1~2}
 \, \cdot \,
{1 \over \vev{1~2}[1~2]}  \, \cdot \,
\frac{\vev{4~1}[2~1]\vev{3~4}}{\vev{4~5}\vev{5~1}[2~1]}
\,=\
\frac{\vev{3~4}\vev{4~1}}{\vev{4~5}\vev{5~1}[1~2]}
 \ .
}

3. The third  contribution \eqref{fe3two}, is
\EQ{ \label{3three3}
\vev{2~3}
 \, \cdot \,
{1 \over \vev{2~3}[2~3]}  \, \cdot \,
\frac{\vev{1~4}^2}{\vev{4~5}\vev{5~1}}
\,=\
 \frac{\vev{1~4}^2}{\vev{4~5}\vev{5~1}[2~3]}
\ .
}

We add up the three contributions \eqref{3one1}, \eqref{3two1}, \eqref{3three3}
and using the momentum conservation identities obtain
\be \label{reaan4}
A^{{  {\mathcal  N}}=4}_5(\Lambda_{(1)\,1}^-,\Lambda_{(2)\,2}^-,
\Lambda_{(3)\,3}^-,\Lambda_{(4)\,4}^-,g^+_5)
= \
-\ \frac{[2~5][3~5]}{[1~2][2~3][3~4]} \ .
\ee
which is again the correct answer for this amplitude, as it can be easily seen from
taking a complex conjugation of the corresponding analytic expression.

\section{NMHV (- - -) Amplitudes with Two Fermions}

In this and the following section we restrict to ${\mathcal N}\le 1$ theory.
There is only one type of fermions, $\Lambda^1 = \Lambda.$
We start with the case of one fermion-antifermion pair, $\Lambda^-$, $\Lambda^+$,
and an arbitrary number of gluons, $g$. The amplitude has a schematic form,
$A_n(\L_{m_1}^-,g_{m_2}^-,g_{m_3}^-,\L_k^+),$
and without loss of generality we can have ${m_1}<m_2<m_3.$
With these conventions,
there are three different classes of amplitudes depending on the position
of the $\Lambda_k^+$ fermion relative to $m_1, m_2, m_3$:
\AL{\label{processes1}
A_n(\L_{m_1}^-,g_{m_2}^-,g_{m_3}^-,\L_k^+)\ , \\
\label{processes2}
A_n(\L_{m_1}^-,g_{m_2}^-,\L_k^+,g_{m_3}^-)\ , \\
\label{processes3}
A_n(\L_{m_1}^-,\L_k^+,g_{m_2}^-,g_{m_3}^-) \ .
}
Each of these three amplitudes receives contributions from different
types of scalar diagrams in the CSW approach. In all of these scalar
diagrams there are precisely two MHV vertices connected to each other
by a single scalar propagator \cite{CSW}. We will always arrange these diagrams
in such a way that the MHV vertex on the left has a positive helicity on the
internal line, and the right vertex has a negative helicity. Then, there are three
choices one can make \cite{Kosower} for the pair of negative helicity
particles to enter external lines of the left vertex,
$({m_1},m_2),$ $(m_2,m_3),$ or $(m_3,{m_1}).$  In addition to this, each diagram
in ${  {\mathcal  N}} \le 1$ theory
corresponds to either a gluon exchange, or a fermion exchange.

\begin{figure}[ht]
\label{fig1}
\epsfxsize=13cm   
\epsfbox{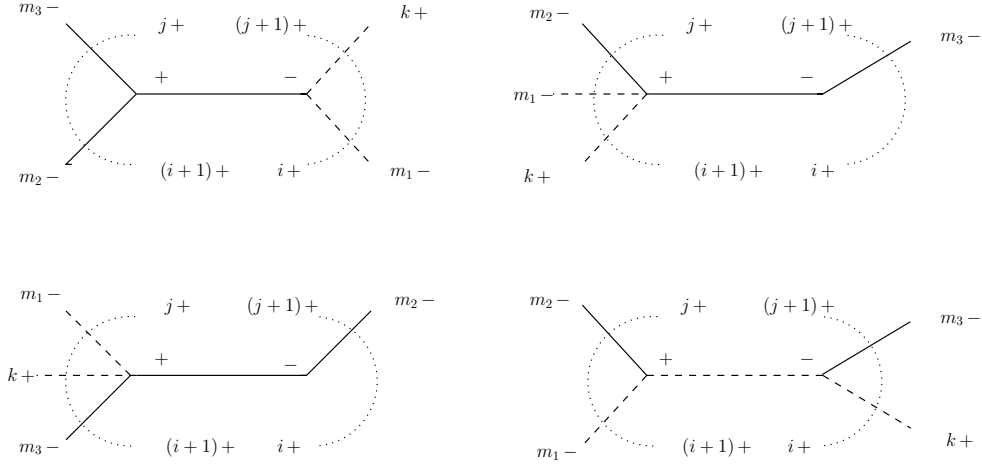}
\caption{Diagrams with MHV vertices contributing to the
amplitude
$A_n(\L_{m_1}^-,g_{m_2}^-,g_{m_3}^-,\L_k^+)$.
Fermions, $\Lambda^+$ and $\Lambda^-,$ are represented by dashed lines and
negative helicity gluons, $g^-,$  by solid lines. Positive helicity gluons
$g^+$ emitted from each vertex are indicated by dotted semicircles with labels
showing the bounding $g^+$ lines in each MHV vertex.}
\end{figure}
The diagrams contributing to the first process \eqref{processes1}
are drawn in Figure 1. There are three gluon exchange diagrams
for all three partitions $(m_2,m_3),$ $({m_1},m_2),$ $(m_3,{m_1}),$
and there is one fermion exchange diagram for the partition $({m_1},m_2).$

It is straightforward, using the expressions for the MHV vertices
\eqref{mpng},\eqref{ndcls},
to write down an analytic expression
for the
first diagram of Figure 1:
\SP{ \label{mmmPPP}
A_n^{(1)} = {1 \over \prod_{l=1}^n\ \vev{l~l+1}}
&\sum_{i={m_1}}^{m_2-1} \sum_{j=m_3}^{k-1}
{-\vev{(i+1,j)~ {m_1}}^3 \ \vev{(i+1,j)~ k}
\over
\vev{i~(i+1,j)}\vev{(i+1,j)~j+1} }  \\
&\times { \vev{i~i+1} \vev{j~j+1}    \over q_{i+1,j}^2}
{\vev{m_2~m_3}^4 \over \vev{(j+1,i)~i+1} \vev{j~(j+1,i)}}
 \ .}
This expression is a direct rendering of the `Feynman rules' for the
scalar graph method \cite{CSW,GK}, followed by factoring out the overall
factor of $(\prod_{l=1}^n\ \vev{l~l+1})^{-1}.$ The objects $(i+1,j)$ and
$(j+1,i)$ appearing on the right hand side of \eqref{mmmPPP}
denote the spinors $\lambda_{i+1,j}$ and $\lambda_{j+1,i}$
corresponding to the off-shell momentum $q_{i+1,j}$
\SP{
&q_{i+1,j} \equiv\ p_{i+1} + p_{i+2}+ \ldots + p_j \ , \
q_{j+1,i} \equiv\ p_{j+1} + p_{j+2}+ \ldots + p_i \ , \\
&q_{i+1,j} + q_{j+1,i} =\ 0  \, \\
&\lambda_{i+1,j\,a} \equiv \ q_{i+1,j\, a\dot a}\, \xi_{\sst \rm Ref}^{\dot a} =\
- \lambda_{j+1,i\,a} \ ,
}
where $\xi_{\sst \rm Ref}^{\dot a}$ is the reference (dotted) spinor \cite{CSW}
as in Eq.~\eqref{ofsh}.
All other spinors $\lambda_i$ are on-shell and
$\vev{i~(j,k)}$ is an abbreviation for a spinor product
$\vev{\lambda_i,\lambda_{jk}}$.

Having the freedom to choose any reference spinor
we will always
choose it to be the spinor of the fermion $\L^-.$
In this section, this is the spinor of $\L_{m_1}^-,$
\be \label{ofshNEW}
\xi_{\sst \rm Ref}^{\dot a}=\ \tilde{\lambda}_{m_1}^{\dot a}
\ .
\ee
We can now re-write
\SP{
&\vev{i~(i+1,j)}\vev{(i+1,j)~j+1}
\vev{(j+1,i)~i+1} \vev{j~(j+1,i)} \\
&=\vev{i^-|{q\!\!\!/}_{i+1,j}| m_1^-}
\vev{j+1^-|{q\!\!\!/}_{i+1,j}| m_1^-}
\vev{i+1^-|{q\!\!\!/}_{i+1,j}| m_1^-}
\vev{j^-|{q\!\!\!/}_{i+1,j}| m_1^-} \ ,}
and define a universal combination,
\SP{ \label{Ddef}
D
=\vev{i^-|{q\!\!\!/}_{i+1,j}| m_1^-}
\vev{j+1^-|{q\!\!\!/}_{i+1,j}| m_1^-}
&\vev{i+1^-|{q\!\!\!/}_{i+1,j}| m_1^-}
\vev{j^-|{q\!\!\!/}_{i+1,j}| m_1^-} \\
&\times\frac{q_{i+1,j}^2}{\vev{i~i+1}\vev{j~j+1}}
}
Note that
Here we introduced the standard Lorentz-invariant
matrix element
$\vev{ i^-|{{p\!\!\!/}\;}_k|j^-} =i^a \,p_{k~a\dot a}\,j^{\dot a}$,
which in terms of the spinor products  is
\be \label{lorinv}
\vev{ i^-|{{p\!\!\!/}\;}_k|j^-} = \
\vev{ i^-|^a \ |k^+}_a \ \vev{k^+|_{\dot a} \ |j^-}^{\dot a} = \
-\vev{i~k} \, [k~j] = \ \vev{i~k} \, [j~k] \ .
\ee

The expression for $A_n^{(1)} $ now becomes:
\SP{ \label{A1}
A_n^{(1)} = {-1 \over \prod_{l=1}^n\ \vev{l~l+1}}
\sum_{i=m_1}^{m_2-1} \sum_{j=m_3}^{k-1}{\vev{m_1^-|{q\!\!\!/}_{i+1,j}|m_1^-}^3
\vev{k^-|{q\!\!\!/}_{i+1,j}| m_1^-}\vev{m_2~m_3}^4
\over
D }\
\ .}

For the second diagram of Figure 1, we have
\SP{ \label{A2}
A_n^{(2)} = {-1 \over \prod_{l=1}^n\ \vev{l~l+1}}
\sum_{i=m_3}^{k-1} \sum_{j=m_2}^{m_3-1}{\vev{m_3^-|{q\!\!\!/}_{i+1,j}|m_1^-}^4
\vev{m_2~m_1}^3\vev{m_2~k}
\over
D }\
\ .}
The MHV vertex on the right
in the second diagram in Figure 1
can collapse to a 2-leg vertex. This occurs when $i=m_3$ and $j+1=m_3.$
This vertex is identically zero, since $q_{j+1,i}=p_{m_3}=-q_{i+1,j},$
and
$\vev{m_3~m_3}=0.$ Similar considerations apply in
\eqref{A3}, \eqref{A2'},
\eqref{A4'}, \eqref{A22},
\eqref{A23}, \eqref{A2''} and
\eqref{tA5}.

Expressions corresponding to the third and fourth diagrams in Figure 1 are
\begin{eqnarray}
 \label{A3}
A_n^{(3)} &=& {-1 \over \prod_{l=1}^n\ \vev{l~l+1}}
\sum_{i=m_2}^{m_3-1} \ \sum_{j=m_1}^{m_2-1}{\vev{m_2^-|{q\!\!\!/}_{i+1,j}|m_1^-}^4
\vev{m_3~m_1}^3\vev{m_3~k}
\over
D } \\
A_n^{(4)} &=& {-1 \over \prod_{l=1}^n\ \vev{l~l+1}} \nonumber\\
&\times&\sum_{i=k}^{n+m_1-1} \ \sum_{j=m_2}^{m_3-1}
{\vev{m_3^-|{q\!\!\!/}_{i+1,j}|m_1^-}^3
\vev{m_2^-|{q\!\!\!/}_{i+1,j}|m_1^-}
\vev{m_2~m_1}^3\vev{m_3~k}
\over
D \label{A4}}
\end{eqnarray}
Note that the first sum in \eqref{A4}, $\sum_{i=k}^{n+m_1-1},$ is understood to run in cyclic
order, for example $\sum_{i=4}^{3} = \sum_{i=4,\ldots,n,1,2,3}.$ The same comment will also apply to
similar sums in
Eqs.~\eqref{A1'}, \eqref{A4'}, \eqref{A12}, \eqref{A22} below.

The total amplitude is the sum
of \eqref{A1}, \eqref{A2}, \eqref{A3} and \eqref{A4},
\bea
A_n(\L_{m_1}^-,g_{m_2}^-,g_{m_3}^-,\L_k^+) = \ \sum_{i=1}^{4} A_n^{(i)} \ .
\eea

There are three sources of zeroes in the denominator combination $D$ defined in
\eqref{Ddef}.   First, there are genuine zeroes in, for example,
$\vev{i^-|{q\!\!\!/}_{i+1,j}| m_1^-}$ when $q_{i+1,j}$ is proportional to
$p_i$.  This occurs when $j=i-1$.   Such terms are always associated with
two-leg vertices as discussed above and produce zeroes
in the numerator.   In fact, the number of zeroes in the numerator always
exceeds the number of zeroes in the denominator and this contribution vanishes.
Second, there are zeroes associated with three-point vertices when,  for example, $i=m_2$ and
$q_{i+1,j}= p_{m_2}+p_{m_1}$ so that
$\vev{m_2^-|{m\!\!\!/}_{1}+{m\!\!\!/}_{2}| m_1^-} = 0$.
As discussed in Sec.~2, there is always a compensating factor in the numerator.
Such terms give a finite contribution (see \eqref{simpl2}).
Third, there are accidental zeroes when $q_{i+1,j}$ happens to be
a linear combination of $p_i$ and $p_{m_1}$.   For general phase space points
this is not the case.
However, at certain phase space points,  the Gram determinant of
$p_i$, $p_{m_1}$ and $q_{i+1,j}$ does vanish.  This produces an apparent
singularity in individual terms in \eqref{A1}--\eqref{A4} which
cancels when all contributions are taken into account.
This cancellation can be achieved numerically or
straightforwardly eliminated using standard spinor techniques~\cite{Kosower}.

For the special case of coincident negative helicities, $m_1=1$, $m_2 = 2$, $m_3 = 3$,
the double sums in Eqs.~\eqref{A1}--\eqref{A4} collapse to single sums.
Furthermore,
we see that the contribution from \eqref{A3} vanishes due to momentum
conservation, $q_{2,1}=0$.  The remaining three
terms agree with the result presented in Eq.~(3.6) of Ref.~\cite{GK}.

We now consider the second amplitude,
Eq.~\eqref{processes2}. The scalar graph diagrams
are shown in Figure 2.
\begin{figure}[ht]
\label{fig2}
\epsfxsize=13cm   
\epsfbox{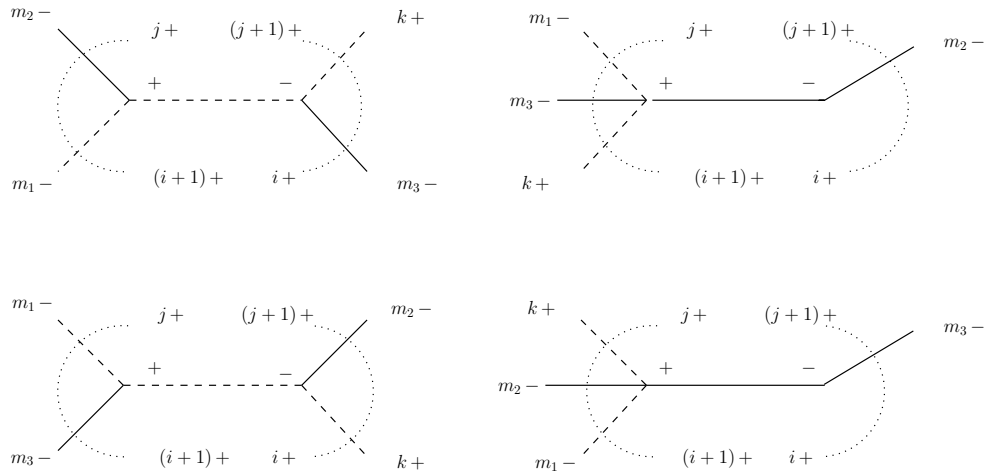}
\caption{Diagrams with MHV vertices contributing to the
amplitude
$A_n(\L_{m_1}^-,g_{m_2}^-,\L_k^+,g_{m_3}^-)$.}
\end{figure}
There is a fermion exchange
and a gluon exchange diagram for two of the line assignments,
 $(m_1,m_2),$ and $(m_3,m_1),$  and none for the remaining assignment $(m_2,m_3).$
These four diagrams result in:
\begin{eqnarray}
A_n^{(1)'} &=& {1 \over \prod_{l=1}^{n}\ \vev{l~l+1}} \nonumber\\
 \label{A1'} &\times&
\sum_{i=m_3}^{n+m_1-1} \ \sum_{j=m_2}^{k-1}{\vev{m_3^-|{q\!\!\!/}_{i+1,j}|m_1^-}^3
\vev{m_2^-|{q\!\!\!/}_{i+1,j}|m_1^-}
\vev{m_2~m_1}^3\vev{m_3~k}
\over
D } \nonumber
 \\
 \label{A2'}
A_n^{(2)'} &=& {1 \over \prod_{l=1}^n\ \vev{l~l+1}}
\sum_{i=m_2}^{k-1}\ \sum_{j=m_1}^{m_2-1}{\vev{m_2^-|{q\!\!\!/}_{i+1,j}|m_1^-}^4
\vev{m_3~m_1}^3\vev{m_3~k}
\over
D }\nonumber
 \\
A_n^{(3)'} &=& {1 \over \prod_{l=1}^n\ \vev{l~l+1}} \nonumber\\
 \label{A3'} &\times&
\sum_{i=k}^{m_3-1} \ \sum_{j=m_1}^{m_2-1}{\vev{m_2^-|{q\!\!\!/}_{i+1,j}|m_1^-}^3
\vev{m_3^-|{q\!\!\!/}_{i+1,j}|m_1^-}
\vev{m_3~m_1}^3\vev{m_2~k}
\over
D } \nonumber
 \\
 \label{A4'}
A_n^{(4)'} &=& {-1 \over \prod_{l=1}^n\ \vev{l~l+1}}
\sum_{i=m_3}^{n+m_1-1}\ \sum_{j=k}^{m_3-1}{\vev{m_3^-|{q\!\!\!/}_{i+1,j}|m_1^-}^4
\vev{m_2~m_1}^3\vev{m_2~k}
\over
D } \nonumber
\end{eqnarray}
and the final answer for \eqref{processes2} is,
\bea
A_n(\L_{m_1}^-,g_{m_2}^-,\L_k^+,g_{m_3}^-) = \ \sum_{i=1}^{4}A_n^{(i)'} \ .
\eea

Finally, we give the result for \eqref{processes3}. The corresponding diagrams
are drawn in Figure 3.
\begin{figure}[ht]
\label{fig3}
\epsfxsize=13cm   
\epsfbox{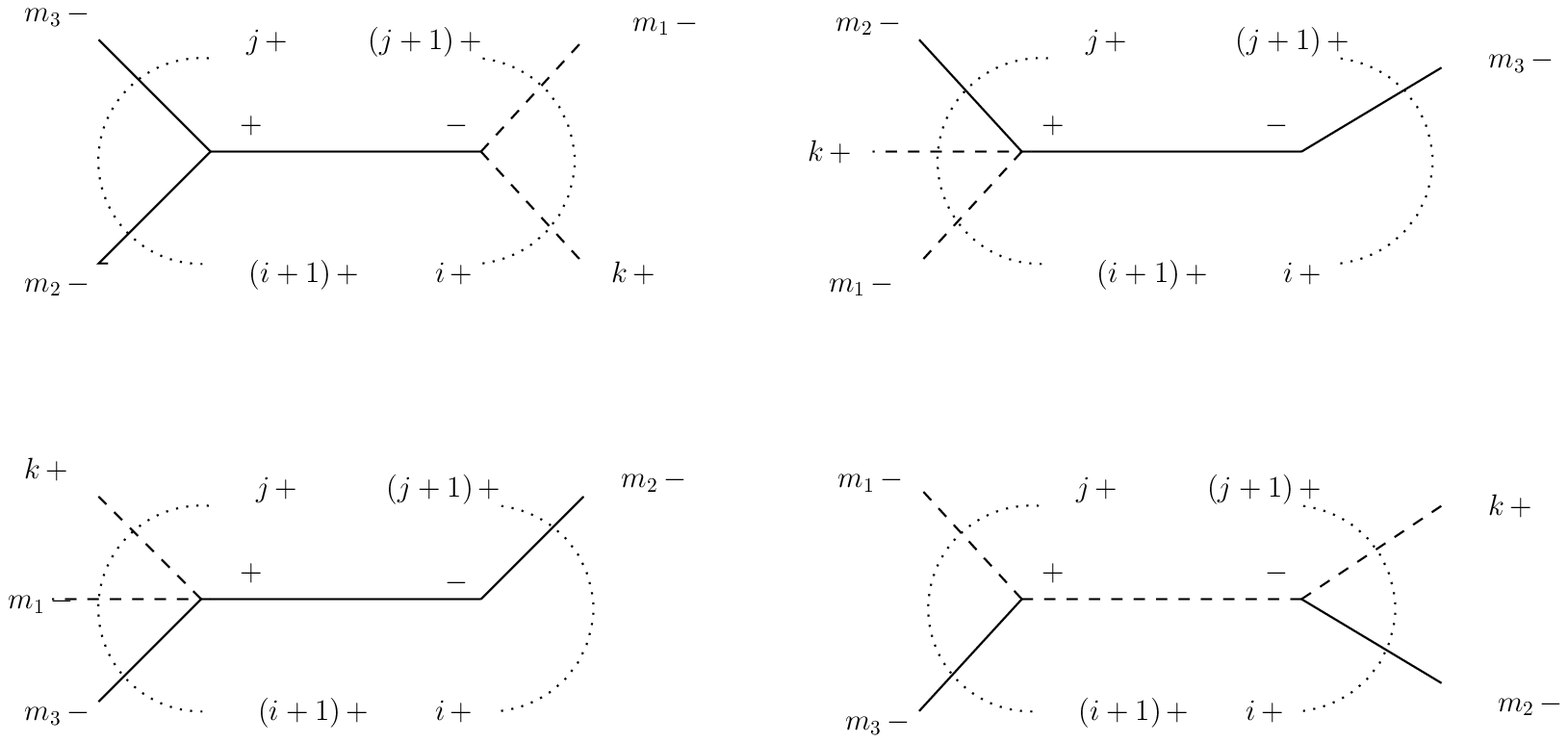}
\caption{Diagrams with MHV vertices contributing to the
amplitude
$A_n(\L_{m_1}^-,\L_k^+,g_{m_2}^-,g_{m_3}^-)$.}
\end{figure}

We find
\begin{eqnarray}
\label{A12}
A_n^{(1)''} &=& {1 \over \prod_{l=1}^n\ \vev{l~l+1}}
\sum_{i=k}^{m_2-1} \sum_{j=m_3}^{n+m_1-1}{\vev{m_1^-|{q\!\!\!/}_{i+1,j}|m_1^-}^3
\vev{k^-|{q\!\!\!/}_{i+1,j}|m_1^-}\vev{m_2~m_3}^4
\over
D }\nonumber
 \\
 \label{A22}
A_n^{(2)''} &=& {1 \over \prod_{l=1}^n\ \vev{l~l+1}}
\sum_{i=m_3}^{n+m_1-1} \sum_{j=m_2}^{m_3-1}{\vev{m_3^-|{q\!\!\!/}_{i+1,j}|m_1^-}^4
\vev{m_2~m_1}^3\vev{m_2~k}
\over
D }\nonumber
 \\
 \label{A23}
A_n^{(3)''} &=& {1 \over \prod_{l=1}^n\ \vev{l~l+1}}
\sum_{i=m_2}^{m_3-1} \sum_{j=k}^{m_2-1}{\vev{m_2^-|{q\!\!\!/}_{i+1,j}|m_1^-}^4
\vev{m_3~m_1}^3\vev{m_3~k}
\over
D } \nonumber\\
A_n^{(4)''} &=& {-1 \over \prod_{l=1}^n\ \vev{l~l+1}} \nonumber\\
\label{A24}
&\times& \sum_{i=m_2}^{m_3-1} \sum_{j=m_1}^{k-1}{\vev{m_2^-|{q\!\!\!/}_{i+1,j}|m_1^-}^3
\vev{m_3^-|{q\!\!\!/}_{i+1,j}|m_1^-}
\vev{m_3~m_1}^3\vev{m_2~k}
\over
D } \nonumber
\end{eqnarray}
As before, the full amplitude is given by the sum of contributions,
\bea
A_n(\L_{m_1}^-,\L_k^+,g_{m_2}^-,g_{m_3}^-) = \ \sum_{i=1}^{4}A_n^{(i)''} \ .
\eea

\section{NMHV (- - -) Amplitudes with Four  Fermions}

We now consider the amplitudes with 2 fermion-antifermion lines.
In what follows, without loss of generality we will choose the negative helicity gluon to be the
first particle. With this convention, we can write the six
inequivalent amplitudes as:
\AL{ \label{proc1}
A_n(g_1^-,\L_{m_2}^-,\L_{m_3}^-,\L_{m_p}^+,\L_{m_q}^+)\ , \\
\label{proc2}
A_n(g_1^-,\L_{m_2}^-,\L_{m_p}^+,\L_{m_3}^-,\L_{m_q}^+)\ ,\\
\label{proc3}
A_n(g_1^-,\L_{m_2}^-,\L_{m_p}^+,\L_{m_q}^+,\L_{m_3}^-)\ ,\\
\label{proc4}
A_n(g_1^-,\L_{m_p}^+,\L_{m_2}^-,\L_{m_3}^-,\L_{m_q}^+)\ ,\\
\label{proc5}
A_n(g_1^-,\L_{m_p}^+,\L_{m_2}^-,\L_{m_q}^+,\L_{m_3}^-)\ ,\\
\label{proc6}
A_n(g_1^-,\L_{m_p}^+,\L_{m_q}^+,\L_{m_2}^-,\L_{m_3}^-)\ .
}
\begin{figure}[t]
\label{fig4}
\epsfxsize=13cm   
\epsfbox{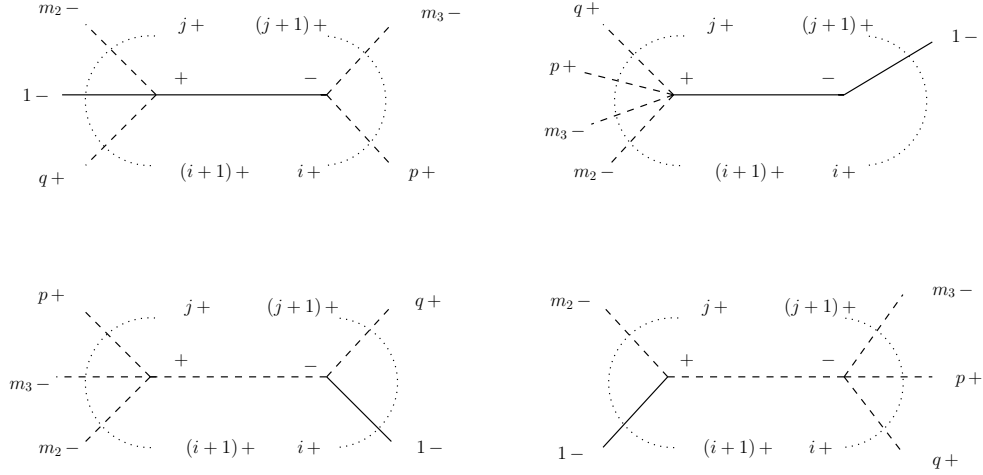}
\caption{Tree diagrams with MHV vertices contributing to the
four fermion amplitude
$A_n(g_1^-,\L_{m_2}^-,\L_{m_3}^-,\L_{m_p}^+,\L_{m_q}^+)$.}
\end{figure}

The calculation of the amplitudes of \eqref{proc1}-\eqref{proc6} is straightforward
\cite{GGK}.
The diagrams contributing to the first process are shown in Figure~4.
It should be noted that not all the amplitudes in \eqref{proc1}-\eqref{proc6} receive
contributions from the same number of diagrams. For example, there are
four diagrams   for the  process of \eqref{proc1} while there are
six for that of \eqref{proc2}.
In order to avoid vanishing denominators, one can choose
the reference spinor to be $\tilde{\eta}=\tilde{\l}_{m_2}$.
With this choice the result can be written as:
\begin{eqnarray}
 \label{A1''}
\tilde{A}_n^{(1)} &=& {1 \over \prod_{l=1}^n\ \vev{l~l+1}}
\sum_{i=p}^{q-1} \sum_{j=m_2}^{m_3-1}{\vev{m_3^-|{q\!\!\!/}_{i+1,j}|m_2^-}^3
\vev{p^-|{q\!\!\!/}_{i+1,j}|m_2^-}
\vev{1~m_2}^3\vev{1~q}
\over
D }\nonumber
 \\
\label{A2''}
\tilde{A}_n^{(2)} &=& {-1 \over \prod_{l=1}^n\ \vev{l~l+1}}
\sum_{i=1}^{m_2-1} \sum_{j=q}^{n}{\vev{1^-|{q\!\!\!/}_{i+1,j}|m_2^-}^4
\vev{m_2~m_3}^3\vev{p~q}
\over
D }\nonumber
  \\
 \label{A3''}
\tilde{A}_n^{(3)} &=& {1 \over \prod_{l=1}^n\ \vev{l~l+1}}
\sum_{i=1}^{m_2-1} \sum_{j=p}^{q-1}{\vev{1^-|{q\!\!\!/}_{i+1,j}|m_2^-}^3
\vev{p^-|{q\!\!\!/}_{i+1,j}|m_2^-}
\vev{m2~m_3}^3\vev{1~q}
\over
D }\nonumber
 \\
\label{A4''}
\tilde{A}_n^{(4)} &=& {-1 \over \prod_{l=1}^n\ \vev{l~l+1}}
\sum_{i=q}^{n} \sum_{j=m_2}^{m_3-1}{\vev{m_3^-|{q\!\!\!/}_{i+1,j}|m_2^-}^3
\vev{1^-|{q\!\!\!/}_{i+1,j}|m_2^-}
\vev{1~m_2}^3\vev{p~q}
\over
D }\nonumber
\end{eqnarray}
As before the final result is the sum
\begin{equation}
A_n(g_1^-,\L_{m_2}^-,\L_{m_3}^-,\L_{m_p}^+,\L_{m_q}^+)=
\sum_{i=1}^4 \tilde{A}_n^{(i)}\ .
\end{equation}
Once again, for the case of coincident negative helicities, $m_2 = 2$, $m_3 =
3$, the double sums collapse to single summations and we recover
the results given in Ref.~\cite{GK}.
\begin{figure}[t]
\label{fig5}
\epsfxsize=13cm   
\epsfbox{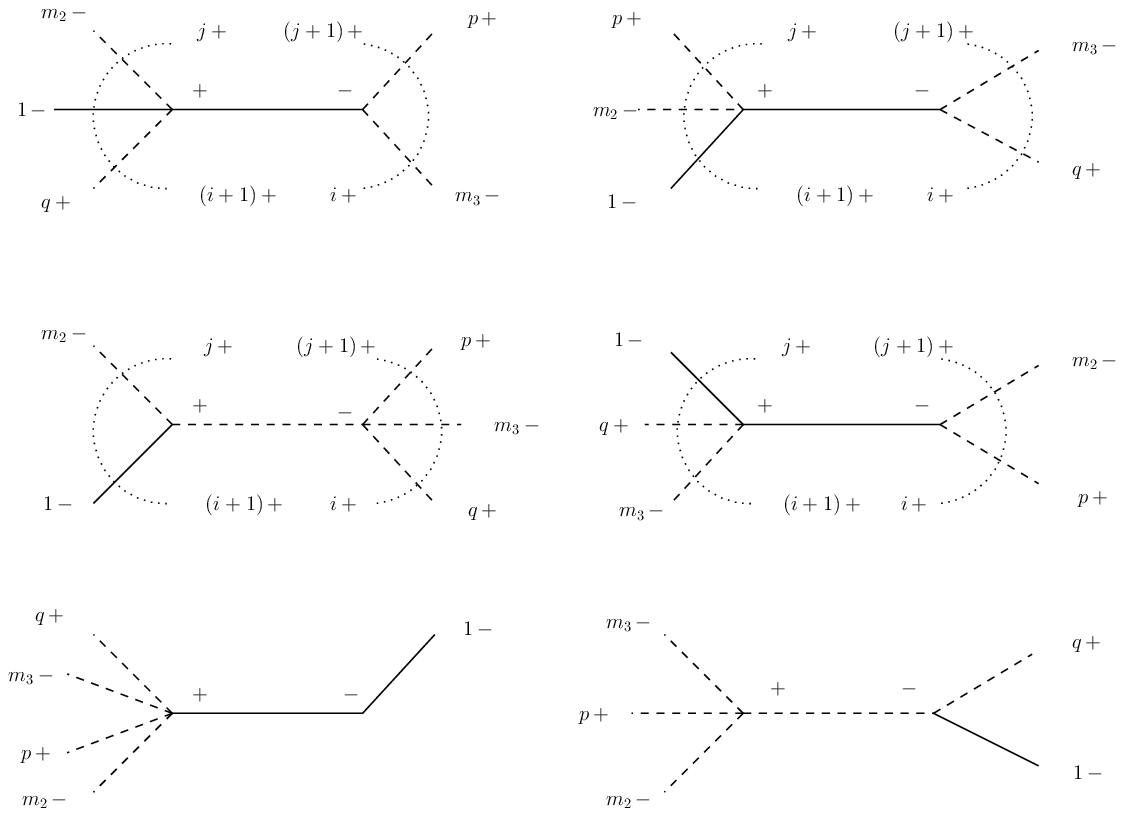}
\caption{Tree diagrams with MHV vertices contributing to the
four fermion amplitude $A_n(g_1^-,\L_{m_2}^-,\L_{m_p}^+,\L_{m_3}^-,\L_{m_q}^+)$.}
\end{figure}
As a last example we write down the expression for the amplitude
of \eqref{proc2}. The corresponding diagrams are shown in Figure 5.
We find,
\begin{eqnarray}
 \label{tA1}
\tilde{A}_n^{(1)'} &=& {-1 \over \prod_{l=1}^n\ \vev{l~l+1}}
\sum_{i=m_3}^{q-1} \sum_{j=m_2}^{p-1}{\vev{m_3^-|{q\!\!\!/}_{i+1,j}|m_2^-}^3
\vev{p^-|{q\!\!\!/}_{i+1,j}|m_2^-}
\vev{1~m_2}^3\vev{1~q}
\over
D }\nonumber
\\
\label{tA2}
\tilde{A}_n^{(2)'} &=& {1 \over \prod_{l=1}^n\ \vev{l~l+1}}
\sum_{i=q}^{n} \sum_{j=p}^{m_3-1}{\vev{m_3^-|{q\!\!\!/}_{i+1,j}|m_2^-}^3
\vev{q^-|{q\!\!\!/}_{i+1,j}|m_2^-}
\vev{1~m_2}^3\vev{1~p}
\over
D }\nonumber
\\
\label{tA3}
\tilde{A}_n^{(3)'} &=& {1 \over \prod_{l=1}^n\ \vev{l~l+1}}
\sum_{i=q}^{n} \sum_{j=m_2}^{p-1}{\vev{m_3^-|{q\!\!\!/}_{i+1,j}|m_2^-}^3
\vev{1^-|{q\!\!\!/}_{i+1,j}|m_2^-}
\vev{1~m_2}^3\vev{p~q}
\over
D }\nonumber
\\
\label{tA4}
\tilde{A}_n^{(4)'} &=& {1 \over \prod_{l=1}^n\ \vev{l~l+1}}
\sum_{i=p}^{m_3-1} \sum_{j=1}^{m_2-1}{\vev{m_2^-|{q\!\!\!/}_{i+1,j}|m_2^-}^3
\vev{p^-|{q\!\!\!/}_{i+1,j}|m_2^-}
\vev{1~m_3}^3\vev{1~q}
\over
D }\nonumber
\\
\label{tA5}
\tilde{A}_n^{(5)'} &=& {1 \over \prod_{l=1}^n\ \vev{l~l+1}}
\sum_{i=1}^{m_2-1} \sum_{j=q}^{n}{\vev{1^-|{q\!\!\!/}_{i+1,j}|m_2^-}^4
\vev{m_2~m_3}^3\vev{p~q}
\over
D }\nonumber
\\
\label{tA6}
\tilde{A}_n^{(6)'} &=& {-1 \over \prod_{l=1}^n\ \vev{l~l+1}}
\sum_{i=1}^{m_2-1} \sum_{j=m_3}^{q-1}{\vev{1^-|{q\!\!\!/}_{i+1,j}|m_2^-}^3
\vev{p^-|{q\!\!\!/}_{i+1,j}|m_2^-}
\vev{m_2~m_3}^3\vev{1~q}
\over
D }\nonumber
\end{eqnarray}
And the full amplitude is
\begin{equation}
A_n(g_1^-,\L_{m_2}^-,\L_{m_p}^+,\L_{m_3}^-,\L_{m_q}^+)=
\sum_{i=1}^6 \tilde{A}_n^{(i)'}\ .
\end{equation}

We close this section by listing the inequivalent
NMHV amplitudes with three fermion--antifermion pairs.
There are ten such amplitudes since choosing the first particle
to be a negative helicity fermion we are left with five fermions
(two of which have negative helicity and three positive) which
should be distributed in all possible ways among themselves, and, in
addition there are $(n-6)$ positive helicity gluons.
Thus the number of different possible ways is $5!$.
However, the order of  the particles of the same helicity
is immaterial (since one can always choose $m_2\leq m_3$
and  $m_p\leq m_q \leq m_r)$. This means that we have to divide
 $5!$ by  $3!$ (for the positive helicity fermions) and by  $2!$
(for the negative helicity fermions.)
Thus there are ten different  fermion amplitudes. These
are listed below:
\SP{
A_n(\L_1^-,\L_{m_2}^-,\L_{m_3}^-,\L_{m_p}^+,\L_{m_q}^+,\L_{m_r}^+)\ , \quad
A_n(\L_1^-,\L_{m_2}^-,\L_{m_p}^+,\L_{m_3}^-,\L_{m_q}^+,\L_{m_r}^+)\ ,\\
A_n(\L_1^-,\L_{m_2}^-,\L_{m_p}^+,\L_{m_q}^+,\L_{m_3}^-,\L_{m_r}^+)\ , \quad
A_n(\L_1^-,\L_{m_p}^+,\L_{m_2}^-,\L_{m_3}^-,\L_{m_q}^+,\L_{m_r}^+)\ ,\\
A_n(\L_1^-,\L_{m_p}^+,\L_{m_2}^-,\L_{m_q}^+,\L_{m_3}^-,\L_{m_r}^+)\ , \quad
A_n(\L_1^-,\L_{m_p}^+,\L_{m_q}^+,\L_{m_2}^-,\L_{m_3}^-,\L_{m_r}^+)\ ,  \\
A_n(\L_1^-,\L_{m_p}^+,\L_{m_q}^+,\L_{m_r}^+,\L_{m_2}^-,\L_{m_3}^-)\ ,  \quad
A_n(\L_1^-,\L_{m_p}^+,\L_{m_q}^+,\L_{m_2}^-,\L_{m_r}^+,\L_{m_3}^-)\ , \\
A_n(\L_1^-,\L_{m_p}^+,\L_{m_2}^-,\L_{m_q}^+,\L_{m_r}^+,\L_{m_3}^-)\ , \quad
A_n(\L_1^-,\L_{m_2}^-,\L_{m_p}^+,\L_{m_q}^+,\L_{m_r}^+,\L_{m_3}^-)\ .
}
These amplitudes also present no difficulty, and they can be evaluated in
the same manner as before.

\section{Two analytic supervertices}

\begin{figure}[ht]
\label{fig6}
\epsfxsize=10cm   
\epsfbox{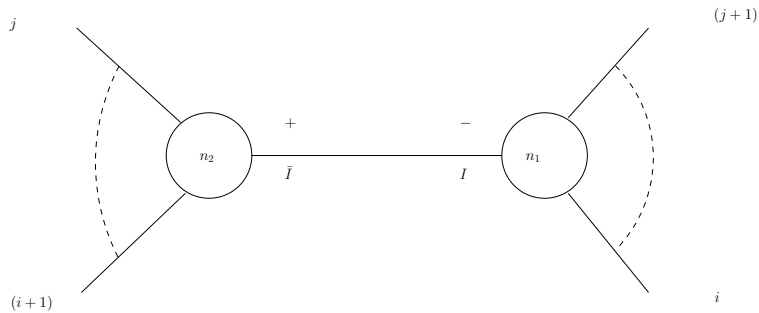}
\caption{Tree diagram with two analytic supervertices.}
\end{figure}
We now consider a diagram with two analytic supervertices
\eqref{nair} connected to one another by a single scalar propagator.
The diagram is depicted in Figure 6. We follow the same conventions
as in the previous sections, and the left vertex has a positive
helicity on the internal line $\bar{I}$, while the right vertex
has a negative helicity on the internal line $I$. The labelling of
the external lines in Figure 6 is also consistent with our conventions.
The right vertex has $n_1$ lines, and the left one has
$n_2$ lines in total, such that resulting amplitude $A_n$ has
$n=n_1+n_2-2$ external lines.
Suppressing summations over the distribution of $n_1$ and $n_2$
between the two vertices, we can write down an expression for the
corresponding amplitude which follows immediately from
\eqref{nair} and Figure 6:
\SP{ \label{itern1}
A_n =\ &{1 \over \prod_{l=1}^n\ \vev{l~l+1}}\
{1 \over q_I^2} \
{\vev{j~j+1} \vev{i~i+1} \over \vev{j~\bar{I}}\vev{\bar{I}~i+1}
\vev{i~I}\vev{I~j+1}} \\
&\times \int \prod_{A=1}^4 d \eta^A_I \
\delta^{(8)} \left(\lambda_{\bar{I}a}\eta_I^A +
\sum_{l_2 \neq \bar{I}}^{n_2} \lambda_{l_2 a}
\eta^A_{l_2} \right) \
\delta^{(8)} \left(\lambda_{Ia}\eta_I^A + \sum_{l_1 \neq I}^{n_1} \lambda_{l_1 a}
\eta^A_{l_1} \right)
 \ .
}
The two delta-functions in \eqref{itern1} come from the two
vertices \eqref{nair}. The summations in the delta-functions arguments run over the
$n_1-1$ external lines for right vertex, and $n_2-1$ external lines for the left one.
The integration over $d^4 \eta_I$ arises in \eqref{itern1} for the
following reason. Two separate (unconnected) vertices in Figure 6
would have $n_1+ n_2$ lines and, hence, $n_1+ n_2$ different
$\eta$'s (and $\lambda$'s). However the $I$ and the $\bar{I}$
lines are connected by the propagator, and there must be only
$n=n_1+n_2-2$ $\eta$-variables left. This is achieved in \eqref{itern1}
by setting
\EQ{ \eta_{\bar{I}}^A = \ \eta_{I}^A \ , }
and integrating over $d^4 \eta_I.$
The off-shell continuation of the internal spinors is defined
as before,
\EQ{
\lambda_{I a} = \ \sum_{l_1 \neq I}^{n_1}\,
p_{l_1\, a \dot{a}}\ \xi_{\sst\rm Ref}^{\dot a} = \ - \lambda_{\bar{I}a} \ .
\label{rlsls}}

We now integrate out four $\eta_I$'s which is made simple by rearranging the arguments of
the delta-functions via $\int \delta(f_2)\delta(f_1) =\int \delta(f_1+f_2)\delta(f_1),$
and noticing that the sum of two arguments, $f_1+f_2,$ does not depend on $\eta_I.$

The final result is
\EQ{\label{finrest}
A_n =\ {1 \over \prod_{l=1}^n\ \vev{l~l+1}}\
\delta^{(8)} \left(\sum_{i=1}^n \lambda_{i a} \eta^A_i \right)\
\prod_{A=1}^4\left(\sum_{l_1 \neq I}^{n_1}\, \vev{I~l_1} \eta^A_{l_1} \right) \
{1\over D} \ ,
}
and $D$ is the same as \eqref{Ddef} used in sections 3 and 4,
\EQ{\label{Pdef}
{1\over D}\ = \ {1 \over q_I^2} \
{\vev{j~j+1} \vev{i~i+1} \over \vev{j~I}\vev{I~i+1}
\vev{i~I}\vev{I~j+1}} \ .
}
There are 12 $\eta$'s in the superamplitude \eqref{finrest},
and the coefficients of the Taylor expansion in $\eta$'s give all the
component amplitudes of degree-12.

\section{One Loop Results}

The next logical step is to extend the formalism to the computation of loop graphs.
The  simplicity and elegant structure of tree level and also loop amplitudes in gauge theory
was quantified in \cite{Witten} by reinterpreting these amplitudes in terms of
a topological string theory with twistor space as a target.

At present we do not
know how to compute SYM loop amplitudes directly from string theory.
It was noted
in \cite{BW} that the currently known topological string models conjectured to be dual to
${\mathcal N}=4$ SYM, at loop level describe SYM coupled to conformal supergravity.
No obvious way was found to decouple supergravitons circulating in the loops.
Also, loop amplitudes in SYM directly in 4 spacetime dimensions
suffer from infrared (IR)
-- soft and collinear -- divergencies.
At tree level there are no integrations over loop momenta and IR
divergencies in the amplitudes can be avoided by selecting a
non-exceptional set of external momenta (i.e the set with none of
the external momenta being collinear or soft).
Hence tree amplitudes can be made IR finite
and it is meaningful to be calculating them
directly in 4D without an explicit IR cutoff.
Loop amplitudes,
however, are always IR divergent and one cannot
choose a set of external momenta which would make an on-shell
loop amplitude finite in 4D. Any successful string computation of loop
amplitudes in gauge theory will have to provide an infrared cutoff,
i.e. a sort of dimensional regularization, but it is not entirely clear at present
how it is encoded in the string with target space $CP^{3|4}$ \cite{GK}.

Having said this, we expect that it is very likely
that twistor space $CP^{3|4}$
will continue to play an important r{\^o}le for understanding
amplitudes at loop level.
The origins of the tree level CSW method \cite{CSW} lied in the unexpected simplicity
of tree-level SYM amplitudes in the helicity basis transformed to twistor space.
Recently
it was shown in \cite{CSW2} that when
 gauge theory amplitudes at 1-loop level are Fourier transformed to twistor space,
their analytic structure again acquires geometric meaning.
It is not know what kind of twistor string theory can generate this geometric structure.

Instead of appealing to string theory, it appears to be more productive
(at present) to calculate loop amplitudes directly in SYM in scalar perturbation theory
of CSW \cite{CSW}.
In order to compute 1-loop amplitudes with the CSW scalar graph method one can choose
two different routes. First is to use the unitarity approach of Bern, Dixon, Dunbar
and Kosower \cite{BDDK} for sewing tree amplitudes to form loops. The CSW method
\cite{CSW} can be used
here to efficiently calculate tree amplitudes
\cite{GK,Kosower,GGK}, as reviewed in sections {\bf 3}, \\
{\bf 7} -- {\bf 9}
above, and 1-loop amplitudes would be obtained from these trees
by sewing them together and using the cut-constructibility method \cite{BDDK}.
This is a promising direction for future study, which cannot fail to lead to new results.

Second approach, is a direct calculation of loop diagrams in the CSW scalar graph
perturbation theory. A priori, there is no proof that the original CSW approach should work
beyond tree level. Moreover, the twistor space motivation (given in section 2 of \cite{CSW2})
of the tree level CSW formalism \cite{CSW}, does not apply directly to loop amplitudes.
Nevertheless, the success of the CSW method at tree level is encouraging enough to
try to apply it a 1-loop level. First such calculations were carried out very recently
by Brandhuber, Spence and Travaglini in \cite{BST}.

The authors of \cite{BST} have calculated 1-loop MHV amplitudes in ${\mathcal  N}=4$ theory
directly using the CSW scalar graph Feynman rules. This is done by taking off-shell
and joining together two external lines from two different vertices
in Figure 6, thus obtaining a 1-loop
MHV diagram with two tree-level analytic supervertices.
Carrying out the integration over the loop
momentum and summing over all inequivalent 1-loop diagrams
gives the final result for this amplitude. Remarkably, this result
turns out to be in precise agreement with the earlier expression derived in \cite{BDDK},
thus vindicating the CSW method at 1-loop level in the simplest case of
${\mathcal  N}=4$ theory and for MHV loop diagrams.

The calculation of \cite{BST} is facilitated by finding a particularly convenient
representation for the integral over the loop momentum. An off-shell loop momentum
$L^\mu$ can be represented as a linear combination of an on-shell momentum
$l^\mu$ and the reference momentum $\xi_{\sst\rm Ref}^{\mu}$ which is also on-shell
\cite{BBK,Kosower}:
\be
L^\mu \, = \, l^{\mu}\, +\, z \,\xi_{\sst\rm Ref}^{\mu} \ , \quad
l^2\,=\,0 \ , \quad \xi_{\sst\rm Ref}^{2}\, = \, 0 \ ,
\ee
and $z$ is a real number,
\be
z\,=\, {L^2 \over 2 (L\,\xi_{\sst\rm Ref})} \ .
\ee
We now write the on-shell vectors $l$ and $\xi_{\sst\rm Ref}$ in terms
of spinors as
$l_{a \dot{a}}=l_a \tilde{l}_{\dot{a}}$ and
$\xi_{\sst\rm Ref}^{\dot{a}a}=
\tilde{\xi}_{\sst\rm Ref}^{\dot{a}} \xi_{\sst\rm Ref}^a $, and find that
\AL{
\label{aaaa}
l_a \ = \ {L_{a \dot{a}}\, \tilde{\xi}_{\sst\rm Ref}^{\dot{a}} \over
[\tilde{l}~\tilde{\xi}_{\sst\rm Ref}]} \ => \
L_{a \dot{a}}\, \tilde{\xi}_{\sst\rm Ref}^{\dot{a}} \ , \\
\label{bbbb}
\tilde{l}_{\dot{a}} \ = \ {\xi_{\sst\rm Ref}^a \,L_{a \dot{a}} \over
\vev{l~\xi_{\sst\rm Ref}}} \ => \ \xi_{\sst\rm Ref}^a \,L_{a \dot{a}} \ .
}
Denominators on the right hand sides of \eqref{aaaa} and \eqref{bbbb} are
dropped because the final expressions for the amplitude are homogeneous in
variables $l$ and $\tilde{l}$.

Note that equation \eqref{aaaa} is identical to the off-shell continuation
prescription \eqref{ofsh} used so far. Hence the off-shell continuation
of external legs for loop amplitudes is precisely the same as at tree level.

The integration over the loop momentum can now be represented in a particularly useful
form \cite{BST} in terms of $z$ and the on-shell spinors $l$, $\tilde{l}$:
\be
{d^4 L \over L^2} \ = \ {dz \over z} \, d \mu(l,\tilde{l}) \ ,
\ee
where $d \mu(l,\tilde{l})$ is the Nair's measure \cite{Nair},
\be
d \mu(l,\tilde{l}) \ = \ \vev{l~dl}\, d^2 \tilde{l} \, - \,
[\tilde{l}~d\tilde{l}]\, d^2 l \ .
\ee
This representation of the integration measure over the loop momentum
in terms of on-shell spinors and the $z$-variable allows a straightforward
evaluation of the loop integral in \cite{BST}.
Another useful property \cite{BST} of the Nair's measure $d \mu(l,\tilde{l})$ is
that it is
equal to the Lorentz-invariant space measure for a massless particle,
\be
d \mu(l,\tilde{l}) \ \sim \ d^4 l \, \delta^{(+)}(l^2) \ .
\ee
This fact makes a remarkable connection between the direct evaluation
of loop integrals and the unitarity approach of \cite{BDDK}.

It will be very interesting to
extend the results of \cite{BST} and to see
if and how the CSW formalism will work in general settings, i.e. at 1-loop and beyond,
for ${\mathcal  N}\le 4$ supersymmetry, and for non-MHV amplitudes.

\section{Conclusions}

We summarize by returning to the questions listed in the introduction.

First we consider tree level amplitudes.
\begin{enumerate}
\item{} The CSW method works in pure gauge sector and in a supersymmetric theory.
This was discussed in the introduction and it
follows from considerations in section {\bf 2.1} and calculations in sections
{\bf 3}, {\bf 6} -- {\bf 8}.
\item{} The method works in a generic supersymmetric gauge theory
with ${\mathcal  N}=1,2,4$ supersymmetries. This follows from calculations in section {\bf 6} and
also from sections {\bf 7} -- {\bf 9}.
\item{} At tree level the method also works in ${\mathcal  N}=0$ theory, such as QCD, and
\item{} It works for finite number of colours, as explained in
the introduction and in section {\bf 2.1}
\item{} The purpose of sections {\bf 7}, {\bf 8} and {\bf 3} and of \cite{GGK}
was to demonstrate that large classes of previously unknown tree amplitudes
with gluons and fermions can now be calculated straightforwardly.
No further helicity-spinor algebra is required to convert the results into
a  numerically usable form. In principle one could use the results presented
here to write a numerical
program for evaluating generic tree-level processes involving fermions and
bosons.

\bigskip

At loop level:
\medskip
\item{} So far the calculations at 1-loop level were carried out only
in ${\mathcal  N}=4$ theory. It is known \cite{BST} that the method works correctly in ${\mathcal  N}=4$ for
MHV amplitudes and at 1-loop.
Given this and the fact that the method was successful at tree level in general settings,
it is likely that it will work for general supersymmetric theories at 1-loop level
and for NMHV loop diagrams.

\item{} There are known difficulties in applying the CSW method ${\mathcal  N}=0$ theories
at 1-loop level, as outlined in section 5.2 of \cite{CSW2}.
At best, the original CSW method needs to be modified by adding additional
off-shell 1-loop vertices
as new building blocks.
\item{} So far the calculations at loop level were performed in the planar limit.
It is not known whether the method can be used to find non-planar contributions.
\end{enumerate}

\newpage
The list of things to do includes:
\begin{itemize}
\item{} Calculate NMHV 1-loop diagrams in ${\mathcal  N}=4$ theory.
\item{} Calculate MHV (and NMHV) amplitudes in ${\mathcal  N}=1$ theory using either
direct loop integrations or by sewing trees.
\item{} Consider modifications of the method for nonsupersymmetric theories
at loop level.
\item{} Include masses.
\item{} Find a twistor space interpretation of the 1-loop calculation
in \cite{BST} and compare it with \cite{CSW2}.
\item{} Search for a string theory calculation of 1-loop amplitudes.
\item{} Understand higher loops, at least in ${\mathcal  N}=4$ SYM.
\end{itemize}

\bigskip

{\bf Acknowledgements}

\medskip
\abstracts{
\noindent This contribution is an extended version of my talk
at `Continuous Advances in QCD 2004' and it follows closely
Refs.~\cite{GK,GGK}.
I thank the organizers for an excellent conference and
George Georgiou and Nigel Glover for an enjoyable collaboration
and for greatly contributing to my understanding of these topics.
I am grateful to
Zvi Bern, Arnd Brandenburg, David Kosower, Misha Shifman, Andrei Smilga, Arkady Vainshtein
and especially to Lance Dixon and Gabriele
Travaglini for useful discussions and comments. I thank the Aspen Center for Physics for
hospitality and workshop participants for many discussions.
This work is supported by a PPARC Senior Fellowship.\\
Dedicated to the memory of Ian Kogan.}




\end{document}